\documentclass[preprint2]{aastex62} %

\hypersetup{linkcolor=blue,citecolor=black,filecolor=cyan,urlcolor=blue}

\usepackage{natbib,layouts}
\newcommand{\1}[1]{\, \mathrm{#1}}

\graphicspath{{./}}

\received{October 20, 2022}
\revised{December 3, 2022}
\accepted{December 28, 2022}
\submitjournal{ApJ}

\shorttitle{Stacked Spectrum with SPT-Selected Sources}
\shortauthors{Reuter et al.}

\begin{document}
\sloppy

\title{The Rest-Frame Submillimeter Spectrum of High Redshift, Dusty, Star-Forming Galaxies from the SPT-SZ Survey}

\correspondingauthor{Cassie Reuter}
\email{creuter@illinois.edu}

\author[0000-0001-7477-1586]{C. Reuter}
\affil{Department of Astronomy, University of Illinois, 1002 West Green St., Urbana, IL 61801, USA}


\author[0000-0003-3256-5615]{J.~S. Spilker}
\affiliation{Department of Physics and Astronomy and George P. and Cynthia Woods Mitchell Institute for Fundamental Physics and Astronomy, Texas A\&M University, 4242 TAMU, College Station, TX 77843-4242, USA}

\author[0000-0001-7192-3871]{J.~D. Vieira}
\affil{Department of Astronomy, University of Illinois, 1002 West Green St., Urbana, IL 61801, USA}
\affil{Department of Physics, University of Illinois, 1110 West Green St., Urbana, IL 61801, USA}
\affil{Center for AstroPhysical Surveys, National Center for Supercomputing Applications, 1205 West Clark Street, Urbana, IL 61801, USA}

\author[0000-0002-2367-1080]{D.~P. Marrone}
\affil{Steward Observatory, University of Arizona, 933 North Cherry Avenue, Tucson, AZ 85721, USA}

\author[0000-0003-4678-3939]{A. Wei{\ss}}
\affil{Max-Planck-Institut f\"{u}r Radioastronomie, Auf dem H\"{u}gel 69, D-53121 Bonn, Germany
}

\author[0000-0002-6290-3198]{M.~Aravena}
\affil{N\'{u}cleo de Astronom{\'i}a de la Facultad de Ingenier\'{i}a y Ciencias, Universidad Diego Portales, Av. Ej\'{e}rcito Libertador 441, Santiago, Chile}

\author[0000-0002-0517-9842]{M.~A.~Archipley}
\affil{Department of Astronomy, University of Illinois, 1002 West Green St., Urbana, IL 61801, USA}
\affil{Center for AstroPhysical Surveys, National Center for Supercomputing Applications, 1205 West Clark Street, Urbana, IL 61801, USA}

\author{S.~C. Chapman}
\affil{Department of Physics and Astronomy, University of British Columbia, 6225 Agricultural Rd., Vancouver, V6T 1Z1, Canada}
\affil{National Research Council, Herzberg Astronomy and Astrophysics, 5071 West Saanich Rd., Victoria, V9E 2E7, Canada}
\affil{Department of Physics and Atmospheric Science, Dalhousie University, Halifax, Nova Scotia, Canada}

\author[0000-0002-0933-8601]{A. Gonzalez}
\affil{Department of Astronomy, University of Florida, 211 Bryant Space Sciences Center, Gainesville, FL 32611, USA}

\author[0000-0002-2554-1837]{T.~R.~Greve}
\affil{Cosmic Dawn Center (DAWN), DTU-Space, Technical University of Denmark, Elektrovej 327, DK-2800 Kgs. Lyngby, Denmark}
\affil{Department of Physics and Astronomy, University College London, Gower Street, London WC1E 6BT, UK}

\author[0000-0003-4073-3236]{C.~C. Hayward}
\affil{Center for Computational Astrophysics, Flatiron Institute, 162 Fifth Avenue, New York, NY, 10010, USA}

\author{R. Hill}
\affil{Department of Physics and Astronomy, University of British Columbia, 6225 Agricultural Rd., Vancouver, V6T 1Z1, Canada}

\author[0000-0002-5386-7076]{S. Jarugula}
\affil{Fermi National Accelerator Laboratory, P.~O.~Box 500, Batavia, IL 60510, USA}

\author[0000-0002-6787-3020]{S.~Kim}
\affil{Department of Astronomy, University of Illinois, 1002 West Green St., Urbana, IL 61801, USA}

\author[0000-0001-6919-1237]{M. Malkan}
\affil{Department of Physics and Astronomy, University of California, Los Angeles, CA 90095-1547, USA}

\author[0000-0001-7946-557X]{K.~A. Phadke}
\affil{Department of Astronomy, University of Illinois, 1002 West Green St., Urbana, IL 61801, USA}
\affil{Center for AstroPhysical Surveys, National Center for Supercomputing Applications, 1205 West Clark Street, Urbana, IL 61801, USA}

\author[0000-0002-2718-9996]{A.~A. Stark}
\affil{Harvard-Smithsonian Center for Astrophysics, 60 Garden Street, Cambridge, MA 02138, USA}

\author[0000-0002-3187-1648]{N. Sulzenauer} 
\affil{Max-Planck-Institut f\"{u}r Radioastronomie, Auf dem H\"{u}gel 69, D-53121 Bonn, Germany
}

\author[0000-0001-7610-5544]{D. Vizgan}
\affil{Department of Astronomy, University of Illinois, 1002 West Green St., Urbana, IL 61801, USA}

\begin{abstract}					%
We present the average rest-frame spectrum of the final catalog of dusty star-forming galaxies (DSFGs) selected from the South Pole Telescope SZ survey (SPT-SZ) and measured with Band 3 of the Atacama Large Millimeter/submillimeter Array (ALMA). 
This work builds on the previous average rest-frame spectrum, given in \citet{spilker14} for the first 22 sources, and is comprised of a total of 78 sources, normalized by their respective apparent dust masses.  The spectrum spans $1.9$$<$z$<$$6.9$ and covers rest-frame frequencies of $240$--$800\1{GHz}$.  
Combining this data with low-J CO observations from the Australia Telescope Compact Array (ATCA), we detect multiple bright line features from $^{12}$CO, $[$CI$]$, and H$_2$O, as well as fainter molecular transitions from $^{13}$CO, HCN, HCO$^+$, HNC, CN, H$_2$O$^+$, and CH.  We use these detections, along with limits from other molecules, to characterize the typical properties of the interstellar medium (ISM) for these high redshift DSFGs.
We are able to divide the large sample into subsets in order to explore how the average spectrum changes with various galaxy properties, such as effective dust temperature.  
We find that systems with hotter dust temperatures exhibit differences in the bright $^{12}$CO emission lines, and contain either warmer and more excited dense gas tracers, or larger dense gas reservoirs.   
These observations will serve as a reference point to studies of the ISM in distant luminous DSFGs (L$_{\mathrm{IR}}$$>$$10^{12}$$\1{L_\odot}$), and will inform studies of chemical evolution before the peak epoch of star formation at $z=2-3$.  
\end{abstract}


\section{Introduction} \label{sec:intro}		         %

In the last two decades, wide-field sub-millimeter (sub-mm) surveys have uncovered a population of very luminous (L$_{\mathrm{IR}} > 10^{12}\1{L_\odot}$), high redshift, dusty star-forming galaxies (DSFGs), with implied star formation rates ranging from hundreds to thousands of solar masses per year (see review by \citealp{casey14}).  
The abundance of DSFGs in the early Universe enables them to contribute significantly to the average star formation rate density ($\sim 80\%$ at $z=2-2.5$ to $\sim 35\%$ at $z=5$; e.g.~\citealp{zavala18,zavala21}).  

The extreme star formation rates of DSFGs imply that they contain vast reservoirs of molecular gas, which is used to form stars.  These reservoirs would be dominated by molecular hydrogen (H$_2$, with M$_{\mathrm{H_2}}\sim 5\times 10^9 - 1.1 \times 10^{11}\1{M_\odot}$ for $z\sim1-3$; e.g.~\citealp{aravena19}).  However, due to the difficulty of detecting molecular hydrogen, other species are commonly used to trace the bulk of the molecular gas in the interstellar medium (ISM).  The rotational ground state of the second most abundant molecule, carbon monoxide ($^{12}$C$^{16}$O $J = 1 \rightarrow 0$ or simply CO(1-0)) is the most common and arguably direct tracer molecular material in the ISM~\citep{carilli13}.  
When the rotational ladder of CO can be measured, it is a powerful probe of the physical conditions of the emitting medium.  However, a number of processes contribute to the excitation of CO, including star formation, shocks, mechanical heating, and potential activity from active galactic nuclei (AGN).  Additionally, the approximate constant of proportionality relating gas mass and CO luminosity can vary by an order of magnitude, as a result of the metallicity and gas conditions of each galaxy~(e.g. \citealp{bolatto13}).  Constraints from other optically thin species, such as $^{13}$CO and C$^{18}$O, could enable a more accurate determination of gas mass, provided the relative abundances of these isotopologues.  However, these species are less abundant and detections of these molecules are limited to a handful of lensed objects (e.g.~\citealp{danielson13}), so their global abundances are relatively unknown for DSFGs.  

Gravitational lensing provides the opportunity to study the gas content of these systems with both greater sensitivity and angular resolution than their unlensed counterparts.  Flux-limited selections of samples discovered with wide-field mm/sub-mm surveys have been shown to efficiently select strongly lensed sources~\citep{vieira13,hezaveh13,spilker16}.  Though lensed DSFGs are relatively rare, ($N<1\1{deg^{-2}}$ for S$_{850\1{\mu m}} > 100\1{mJy}$; e.g. \citealp{negrello07}), hundreds of gravitationally lensed DSFGs have been discovered via wide field surveys with the Atacama Cosmology Telescope (ACT; \citealp{marsden14,su17,gralla20}), \textit{Herschel} (e.g.~\citealp{negrello10, negrello17, nayyeri16, bakx18, urquhart22}), \textit{Planck} (e.g.~\citealp{canameras15, harrington16, berman22}) and the South Pole Telescope (SPT; \citealp{vieira10, vieira13, everett20}).   
Gravitational magnification (typically $\mu \sim 6$; \citealp{spilker16}) enables the observation of lines otherwise too faint to see~\citep{spilker14}.  It also decreases the on-source time required for bright emission lines, facilitating large spectroscopic surveys.  Early CO surveys of high redshift galaxies~\citep{bothwell13, daddi15} studied multiple $^{12}$CO emission lines and provided the first constraints on the CO rotational ladder, also known as the spectral line energy distribution (SLED).  

Telescopes such as the Atacama Large Millimeter/submillimeter Array (ALMA) and the Northern Extended Millimeter Array (NOEMA) have been conducting blind CO surveys targeting the same $^{12}$CO emission lines~\citep{bothwell13, weiss13, strandet16, reuter20, neri20} in order to determine the spectroscopic redshifts of the targeted sources.  The spectroscopic data collected also serves to probe the ISM of both the individual sources and the underlying DSFG distribution.  The creation of composite spectra by combining survey data~\citep{spilker14, boogaard20, birkin21} places constraints on the ``average" DSFG for bright molecular lines such as $^{12}$CO, [CI] and H$_2$O.  With additional sources, the corresponding reduction in noise also enables the detection of much fainter molecular transitions from $^{13}$CO, HCN, HNC, HCO$^+$, and CN at high redshift.  Taken together, the relative strengths of these lines provide first attempts to quantify the relative abundances of various molecules and their brightest isotopologues in DSFGs and provide insights about the interstellar chemistry at high redshift.  

Molecules with high dipole moments, such as HCN, HNC, HCO$^+$ and CN are excited out of their $J=0$ ground state in high density gas ($n>10^4\1{cm^{-3}}$), where molecular clouds decouple from external tidal forces and UV background, allowing them to collapse.  These molecules are therefore thought to be reliable tracers of the dense molecular gas associated with star formation~\citep{gao05b, carilli13}.  Emission lines for the majority of these molecules are typically 1--2 orders of magnitude fainter than the bright CO lines targeted (e.g.~\citealp{jimenez-donaire19}) and the early detections were limited to a handful of highly magnified luminous objects~(e.g. the ``Cloverleaf" quasar, see~\citealp{solomon03} and APM 08279+5255, see~\citealp{wagg05}).  While more recent work has included small samples of highly lensed objects (e.g. a selection of strongly-lensed DSFGs from \citealp{bethermin18}), the number of observations at high redshift is still limited.  However, an increasing number of studies indicate a trend between the dense gas tracer HCN and far-IR luminosity~\citep{gao04a, graciacarpio06, juneau09, garciaburillo12}, demonstrating the potential utility of dense gas tracers to trace molecular gas and potentially star formation.  

Water is one of the most abundant molecules in galaxies, though most of it is frozen onto dust grains in cold environments~\citep{tielens91, hollenbach09}.  In warm environments, such as near stars or AGN, these icy dust grains heat through collisional excitation, sublimating water, and make water a tracer of star formation and AGN activity.  
Due to the high optical depth of its lines, water lines can be as luminous as CO lines in the same frequency range~\citep{vanderwerf11}.  Water has been shown to be directly proportional to the infrared luminosity (L$_{\mathrm{IR}}$; \citealp{omont13, yang13}) and has shown potential to serve as a resolved tracer of star formation~\citep{jarugula19}.  In contrast, a deep absorption feature at the rest-frame frequency of $557\1{GHz}$ due to the H$_2$O(1$_{1,0}$-1$_{0,1}$) transition has been observed for a DSFG at $z\sim 6$~\citep{riechers22}.  Because this absorption line may also show absorption into the CMB, it could indicate the presence of cold water vapor.  

With a large enough number of objects, we can split the sample to see how these molecular tracers are affected by various properties, such as effective dust temperature, IR luminosity, stellar mass, etc.  The excitation of bright $^{12}$CO emission has been shown to correlate well with dust temperature~\citep{rosenberg15}.  A previous attempt to understand the relative abundances of fainter molecular lines with respect to mid-IR was inconclusive~\citep{spilker14}, as the sample size was not large enough to overcome the effects of individual source variations.  

In this paper, we will use the data collected from the ALMA $3\1{mm}$ redshift search published in \citet{weiss13}, \citet{strandet16} and \citet{reuter20}, which are briefly summarized in Sec.~\ref{sec:observations}.  The data are scaled and stacked according to Sec.~\ref{sec:stackmethods} in order to create a composite spectrum, presented in Sec.~\ref{sec:stacked_spec}.  We analyze the average conditions of the ISM for this collection of DSFGs in Sec.~\ref{sec:discussion}, and place these results in context with other works from literature.  Our conclusions are summarized in Sec.~\ref{sec:conclusions}.  

For this paper, we adopt a flat Lambda cold dark matter cosmology, with $\Omega_\Lambda$ = 0.696 and $H_0 = 68.1\1{km \, s^{-1} Mpc^{-1}}$~\citep{planck16cosmo}. 

\section{Data and Methods} \label{sec:observations}	%

\subsection{ALMA $3\1{mm}$ Observations} \label{sec:3mmdata}
This work utilizes data from the SPT-SZ DSFG catalog, and the full details of the target selection and ALMA $3\1{mm}$ observations can be found in \citet{reuter20}.  We will briefly summarize the key points in this section.

The SPT DSFG catalog was selected from the $2500\1{deg^2}$ field of the SPT-Sunyaev-Zel'dovich survey (SPT-SZ; \citealp{vieira10, everett20}) with $\mathrm{S}_{1.4\1{mm}} > 20\1{mJy}$.  In order to refine their positions, the sources are also required to have been detected with the Large Apex BOlometer CAmera (LABOCA; \citealp{siringo09}) with $\mathrm{S}_{870\1{\mu m}} > 25\1{mJy}$, bringing the number of sources in the flux-limited selection to 81.  Over 70$\%$ of the sources in the sample are strongly gravitationally lensed by galaxies~\citep{spilker16}, and with a median magnification factor of $\mu_{870\1{\mu m}} = 5.5$.  The remainder of the sample are either lensed by galaxy clusters or are unlensed protoclusters (e.g. SPT2349-56, see~\citealp{miller18, hill20, wang21}).  The sources span redshifts from $z=1.9-6.9$, apparent IR luminosities from $1.3-29 \times 10^{13}\1{L_\odot}$ (integrated from $8-1000\1{\mu m}$) and dust temperatures from $21-79\1{K}$, with medians at $\langle z \rangle = 3.9 \pm 0.2$, apparent $\langle \mathrm{L_{IR}} \rangle = 8.2 \times 10^{13}\1{\mathrm{L_\odot}}$, and  $\langle \mathrm{T_{dust}} \rangle = 54\1{K}$.  

The ALMA $3\1{mm}$ observations were obtained over several different ALMA Cycles, ranging from Cycle 0 to Cycle 7.\footnote{The data from the following ALMA projects were used in this work: \#2011.0.00957.S, \#2012.1.00844.S, \#2015.1.00504.S, \#2016.1.00672.S and \#2019.1.00486.S.}
The first observations were conducted in ALMA Cycle 0 and spectra were obtained for 26 targets~\citep{weiss13}.  Followup surveys conducted in ALMA Cycles 1, 3, and 4 targeted additional sources, giving a total of 78 spectral scans obtained with ALMA~\citep{strandet16,reuter20}.  While there are 81 sources in the complete SPT DSFG sample, spectroscopic redshifts for three sources (SPT0538-50, SPT0551-48 and SPT2332-53; \citealp{greve12, strandet16}) were obtained with Z-Spec~\citep{bradford09} on the Caltech Submillimeter Observatory and no additional $3\1{mm}$ data were obtained with ALMA.  These sources are therefore not included in the analysis presented in this work.  

Despite the changing capabilities of ALMA during our observations, the observing strategy remained the same for all of the $3\1{mm}$ observations.  The data were obtained by conducting a spectral sweep of ALMA Band 3 in 5 tunings, scanning the range between $84.2$ and $114.9\1{GHz}$.  The brightest sources were targeted in ALMA Cycle 0 and were scanned for roughly 5 minutes on-source, with between 14-17 antennas.  The remaining fainter sources were scanned with both longer integration times and additional antennas in subsequent ALMA Cycles.  Sources scanned in Cycles 1, 3, and 4 were scanned for an average of 10, 6 and 13 minutes on source with between 28-40, 34-41 and 38-46 antennas, respectively.  Finally, three sources did not exhibit any emission lines in their initial $3\1{mm}$ scans, so deeper observations were obtained in ALMA Cycle 7.  These scans observed each target for 45-91 minutes on source with 42-49 antennas.  

Data were processed using the Common Astronomy Software Application (\texttt{CASA}; \citealp{mcmullin07}).  Because integration times and number of antennas increased with each ALMA Cycle, the typical noise levels vary according to the ALMA Cycle and were $1.5\1{mJy \, beam^{-1}}$, $0.91\1{mJy \, beam^{-1}}$, $0.73\1{mJy \, beam^{-1}}$, $0.58\1{mJy \, beam^{-1}}$, and $0.24\1{mJy \, beam^{-1}}$ over a $62.5\1{MHz}$ bandwidth for Cycles 0, 1, 3, 4, and 7, respectively.  The typical RMS values for individual SPT-SZ DSFGs can be found in Table E1 of \citet{reuter20}.  The spectral resolution of the data, $\sim 1.5\1{km/s}$ over a $2\1{GHz}$ baseband, is much higher than a typical galaxy line width.  This allows many channels to be averaged over for each line, increasing significance.  

As previously noted, this work utilizes the $3\1{mm}$ spectra obtained in \citet{weiss13}, \citet{strandet16} and \citet{reuter20}.  These spectra are central to not only this paper, but also for papers currently in preparation. The $3\1{mm}$ line fluxes from mid-J CO lines will be combined with the line fluxes from low-J CO observations with ATCA. Spectral line energy distributions will be fitted for all sources, which will be examined in a following paper that is in preparation. 

\subsection{Stacking Methods} \label{sec:stackmethods}
Because the details of creating a stacked spectrum are similar to those outlined in~\citet{spilker14}, we summarize the key points here and highlight the differences between the two methods.  

The stacked spectrum presented in this paper is a composite of $3\1{mm}$ spectral scans for the 78 sources observed with ALMA.  Calibrated data cubes were created with \texttt{CASA}'s \texttt{TClean} package and the spectra were extracted.  The continuua of the spectra were fitted with a first order polynomial, excluding channels with signal to noise $>3\sigma$, and then subtracted from each spectrum.  The spectra are then re-scaled to a common rest frame ($z=3$; as in Eq. 1 of~\citealp{spilker14}).  The choice to scale to $z = 3$ is intended to represent the typical redshift of DSFGs.  Since surveys of unlensed DSFGs are largely selected at shorter wavelengths, they tend to be at lower redshift ($z\sim2.3\--2.9$; \citealp{koprowski14, simpson14, simpson17, miettinen15, brisbin17, danielson17, michalowski17}), while the median redshift of the SPT-SZ DSFGs is 3.9~\citep{reuter20}.  

Because our sample is known to contain a large number ($>70\%$; \citealp{spilker16}) of gravitationally lensed sources, it is necessary to account for the magnification. Ideally, the magnification would be calculated by constructing gravitational lens models of the sources~\citep{spilker16}.  However, in the absence of a complete set of gravitational lens models, the individual spectra are normalized by their respective apparent dust masses as a proxy for size, then multiplied by the average apparent dust mass ($\langle \mathrm{M}_{\mathrm{dust}}\rangle = 8.7 \times 10^9 \, \mathrm{M}_\odot$).  It should be noted that a radius could also be derived from the same spectral energy density fit (e.g. Eqs. 1 and 2 from~\citealt{weiss07}).  Radius and dust mass are therefore very strongly correlated, with some intrinsic scatter from the simultaneous fitting of dust temperature.  Since normalizing by the apparent values for radius and dust mass give similar results, we choose to normalize by the apparent dust mass throughout this work.  The values for apparent dust mass used throughout this work can be found in Appendix~\ref{ap:weights}.    

\begin{figure}[!h]
	\centering
	\includegraphics[width=\columnwidth]{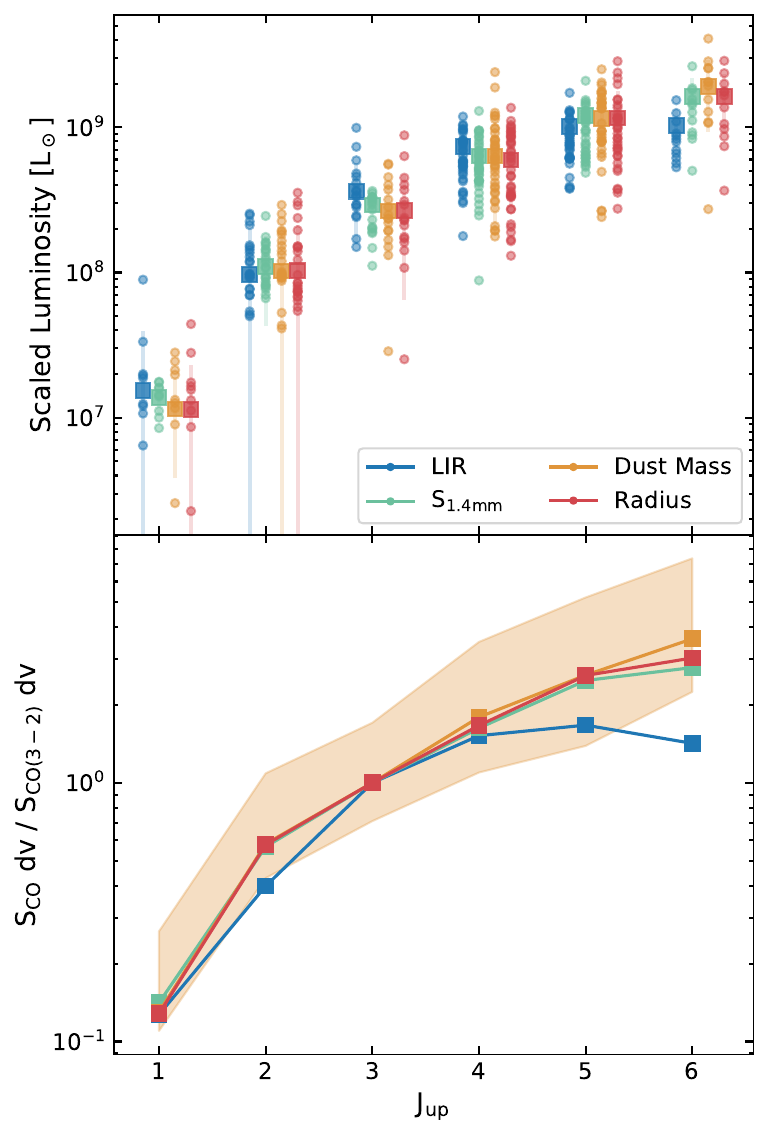}
	\caption{ Spectral power distribution of $^{12}$CO observations.  \textit{Top panel:} The smaller circles represent the CO luminosities of individual sources, while the larger squares represent the averages.  Each individual source is normalized by its respective property, then multiplied by the sample average: apparent L$_{\mathrm{IR}}$ (blue; $\langle \mathrm{L}_\mathrm{IR}\rangle = 9.1 \times 10^{13} \, \mathrm{L}_\odot$), $1.4\1{mm}$ flux (green; $\langle \mathrm{S}_{\mathrm{1.4mm}}\rangle = 18 \, \mathrm{mJy}$), apparent dust mass (yellow; $\langle \mathrm{M}_{\mathrm{dust}}\rangle = 8.7 \times 10^9 \, \mathrm{M}_\odot$) and apparent radius derived from SED fitting (red; $\langle \mathrm{r}\rangle = 7.5\, \mathrm{kpc}$) and scaled according to the average of the sample.  \textit{Bottom panel:}  The average velocity-integrated flux density normalized to average flux density of the J$_{\mathrm{up}}=3$ line, the lowest frequency CO observation obtained from the ALMA $3\1{mm}$ scans.  The 16th and 84th percentiles for the normalization by dust mass are denoted by the shaded region, the others are omitted for clarity.  }
	\label{fig:indvsled}
\end{figure}

A variety of other source properties could also have been used in order to normalize the ALMA spectra, such as apparent L$_{\mathrm{IR}}$, $^{12}$CO line luminosity or scaled $1.4\1{mm}$ SED flux density (as in \citealp{spilker14}).  While the line ratios derived in~\citet{spilker14} were robust within $\sim15\%$ to the choice of normalization, $^{12}$CO line ratios for the complete sample can vary from $\sim 25\%$ to as much as $65\%$.
The large disparity is in part due to the large intrinsic scatter in population statistics shown in Fig.~\ref{fig:indvsled} for a few choices of normalization.  
It should be noted that normalization by IR luminosity will affect the shape of the $^{12}$CO SLED, since the excitation is dependent on IR luminosity (e.g.~\citealp{narayanan08}).  Since the systems at high IR luminosity will be scaled to compare to lower luminosity, this will lead to a lower excitation, as seen in the top panel of Fig.~\ref{fig:indvsled}.  
 
The $1.4\1{mm}$ flux density (rest frame $350\1{\mu m}$) is taken from the modified blackbody models in \citet{reuter20} and represents a compromise between the apparent L$_{\mathrm{IR}}$ and apparent dust mass normalizations.  Because any results derived from a stacked spectrum are dependent on normalization, we acknowledge and emphasize that the stacked spectrum presented in this work is not uniquely defined.  

\begin{figure}[h]
	\centering
	\includegraphics[width=\columnwidth]{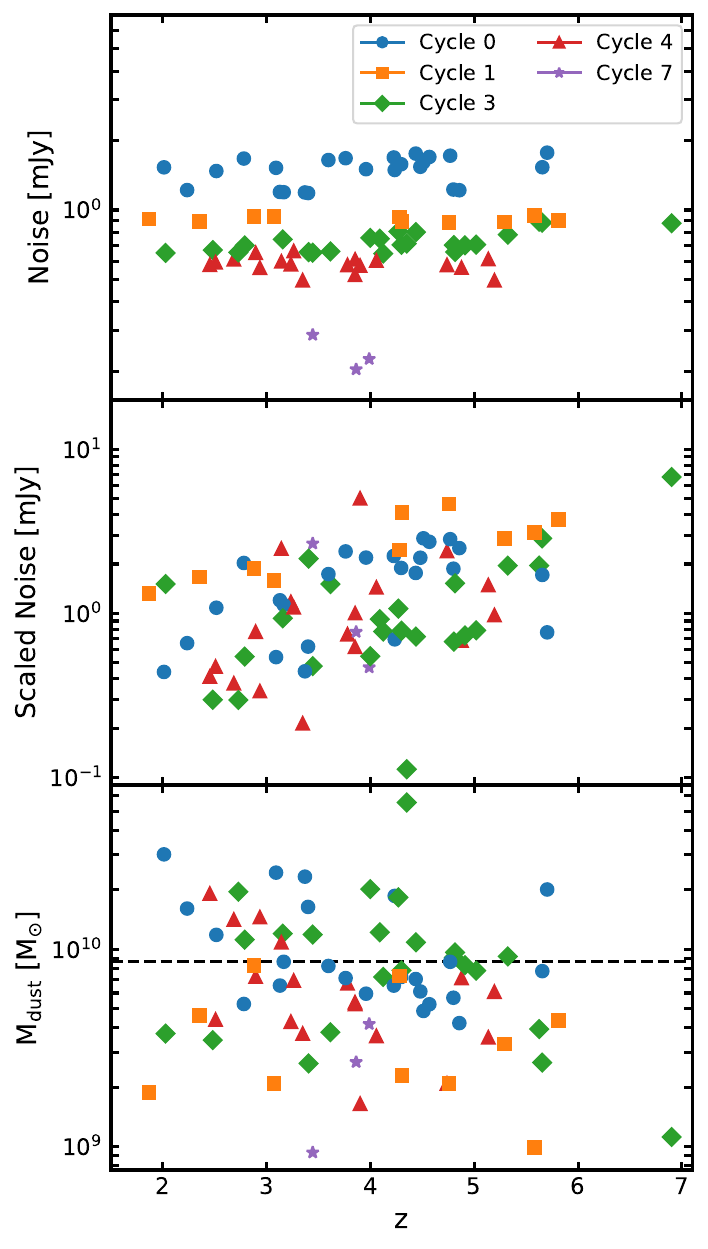}
	\caption{\textit{Top panel:} Noise averaged across the ALMA $3\1{mm}$ window measured in 62.5MHz channels.  Because ALMA added more antennas each Cycle, the noise decreases with progressive ALMA Cycles.  \textit{Bottom panel:} The apparent dust masses for each source, with the sample mean indicated by the horizontal dashed line.  Both the average noise and apparent dust mass are shown in Appendix~\ref{ap:weights}.  \textit{Middle panel:} The weights used in the stacking method, which are the noise from the top panel divided by the fractional difference of the individual dust mass from the median.  All panels are color-coded by the ALMA Cycle number. }
	\label{fig:noise}
\end{figure}

To create the final composite spectrum of the SPT DSFGs, we employ the same methodology as \citet{spilker14}.  That is, the spectrum of each source is interpolated onto a grid spanning the rest frequencies of the ALMA $3\1{mm}$ spectra, $250-770\1{GHz}$.  The grid has a uniform spacing of $500\1{km/s}$, which roughly corresponds to the typical full width half maximum (FWHM) of observed $^{12}$CO transitions.  A weighted average is then performed on each frequency channel, with each contributing source weighted by the inverse standard deviation~\citep{boogaard21}, rather than the inverse variance used in \citet{spilker14}.  Because the data were collected in different ALMA Cycles, the noise of individual spectra over a $62.5\1{MHz}$ bandwidth can vary from the $1.4-2\1{mJy \, beam^{-1}}$ typically observed in Cycle 0 to $0.2-0.3\1{mJy \, beam^{-1}}$ for the three sources re-observed in ALMA Cycle 7 (details in \citealp{reuter20}).  The average $3\1{mm}$ per-channel noise for each source is shown in the top panel of Fig.~\ref{fig:noise} and is also given in Appendix~\ref{ap:weights}.  Though normalization by dust mass (shown in the bottom panel) does mitigate some of the differences in ALMA Cycle, demonstrated by the average weights shown in the middle panel of Fig.~\ref{fig:noise}\footnote{The noise is first scaled to a common redshift ($z=3$), but this step is omitted in Fig.~\ref{fig:noise} in order to show the noise and weights as a function of redshift. }, the choice to weight by inverse standard deviation is a compromise between optimizing signal to noise and accounting for the range of noises due to the varying ALMA Cycles.  

It should be noted that each frequency channel of the contributing $3\1{mm}$ spectrum is weighted individually and the average noise across the $3\1{mm}$ bandwidth is shown in Tab.~\ref{ap:weights}.  Weighting by the inverse variance of noise was used in various composite spectra~\citep{spilker14} in order to optimize for signal to noise.  However, as noted in Sec.~\ref{sec:stackmethods}, if the noise is non-homogeneous (e.g. if there are very different numbers of antennas between observations), weighting by the inverse variance of noise can be sensitive to outliers.  In order to mitigate the presence of outliers, one can choose to weight instead by inverse standard deviation (e.g. \citealp{boogaard21} and this work) or use the median, rather than the average (e.g. \citealp{bothwell13, birkin21}).  A test of the statistical noise properties is given in Appendix~\ref{ap:weights}, and demonstrates the noise is Gaussian across all ALMA Cycles.  

\subsection{$^{12}$CO Line Ratios} 
\begin{figure}[!h]
	\centering
	\includegraphics[width=\columnwidth]{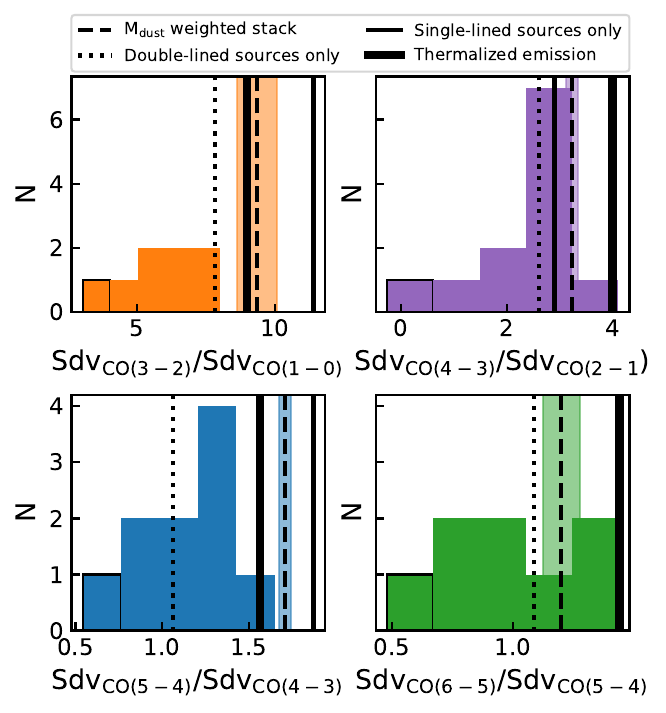}
	\caption{Histograms of $^{12}$CO emission line ratios for sources with two or more observed emission lines, using integrated flux density.  The J$_{\mathrm{up}} = 3-6$ lines were observed with ALMA, while J$_{\mathrm{up}} = 1-2$ lines were observed with ATCA.  The dashed line represents the average line ratio obtained from the composite spectrum, and the thick black line represents the thermalized emission limit. The dotted line represents a stack using the double-lined sources only, while the thin black line corresponds to a stack of the singly-lined sources only.  }
	\label{fig:lineratios}
\end{figure}

For $\sim 32\%$ of the SPT DSFG sample, two or more CO emission lines fall within the $3\1{mm}$ observation window and are detected in the blind survey~\citep{reuter20}.  
Low-J CO lines were obtained from an ATCA CO survey detailed in \citet{aravena16}.  In addition to the published observations (17 sources), data was obtained after publication, giving a total of 33 observations for the J$_{\mathrm{up}} = 1-2$ lines.  Sources that had multiple CO line observations either through ALMA or through the combination of ATCA and ALMA data are so-called ``double-lined" sources.  These sources enable a comparison of the relative strengths of observed $^{12}$CO emission lines within individual systems, shown in Fig.~\ref{fig:lineratios} as histograms, with the optically thick thermalized emission limit represented by the thick black line.  

The $^{12}$CO emission line ratios from individual SPT systems also provide a direct comparison sample to the composite spectrum obtained from stacking, which is represented by the dashed line in Fig.~\ref{fig:lineratios}.  
With the exception of CO(5-4)/CO(4-3), the emission line ratios obtained from the composite spectrum are in agreement with the individual ratios and below the optically thick thermalized emission limit.

While the composite spectrum necessarily contains all of the individual line ratios represented in these histograms, it differs in two key respects: each observation is normalized by dust mass and weighted by the inverse of the noise and also the stack contains many more observations of single emission lines.  This is especially apparent in the CO(5-4)/CO(4-3) panel of Fig.~\ref{fig:lineratios}, where the composite line ratio is above both the individual line ratios and the thermalized emission line.  
One reason for this apparent discrepancy is that because of the fixed observation window, the singly-lined sources being compared are necessarily at different redshifts.  While \citet{reuter20} did not find a statistically significant evolution of dust temperature as a function of redshift, two populations of singly-lined sources being compared likely have different effective dust temperatures and luminosities.  Additionally, some of the faintest and lowest mass sources observed in Cycle 7 (e.g. SPT0112-55, SPT0457-49 and SPT2340-59) are singly-lined sources for this analysis as they contain a single $^{12}$CO emission line, while a high dust mass outlier (SPT0155-62) is a double-lined source.  It is likely that the population of sources is simply not large enough yet to overcome the effects of individual source variations, in addition to the effects of potential redshift evolution of the $^{12}$CO line properties given in the fixed observed frequency range.  

\section{The Stacked Spectrum} \label{sec:stacked_spec}			%

\begin{figure*}[!htb]
	\centering
	\includegraphics[width=\textwidth]{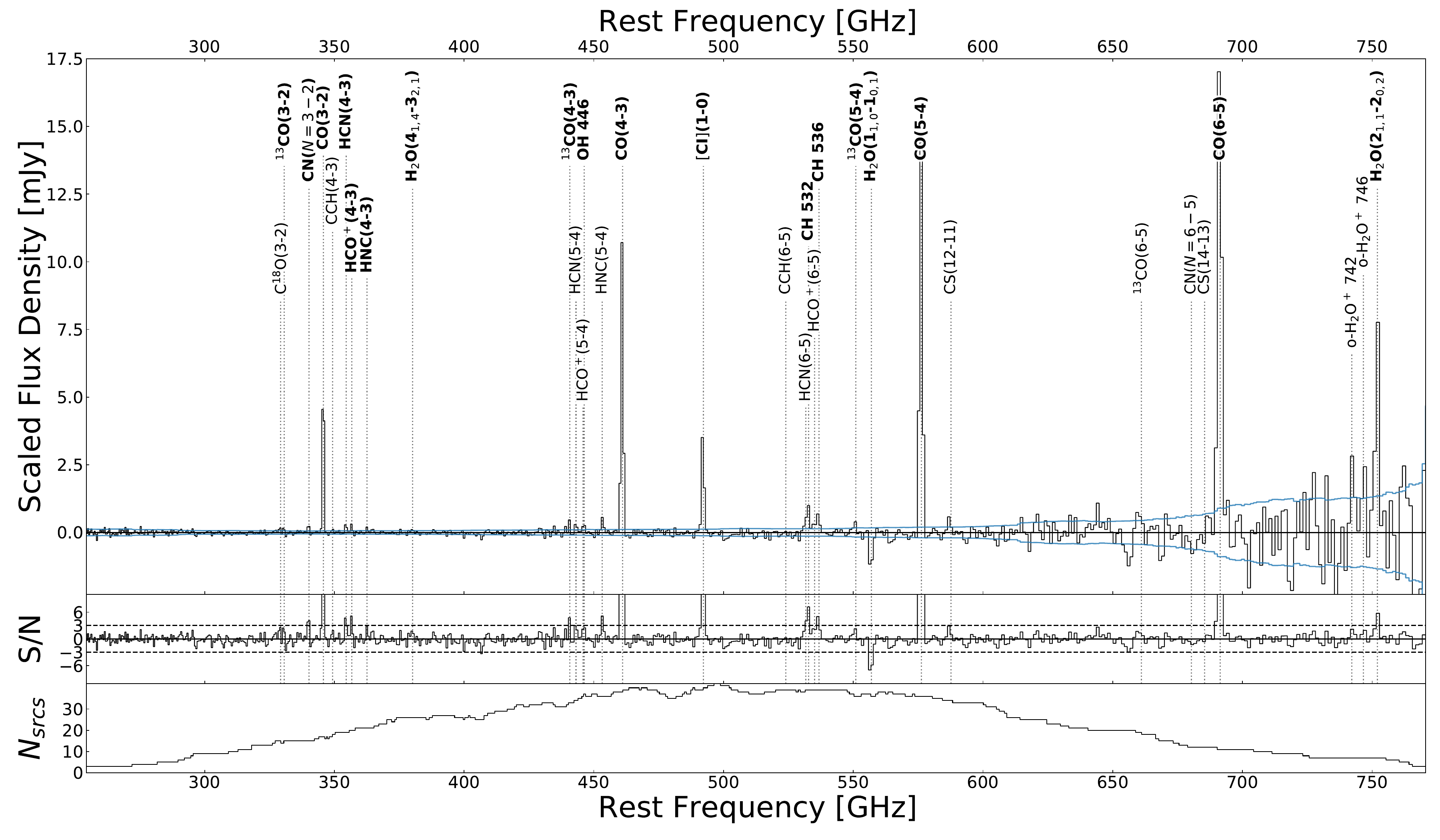}
	\caption{Composite continuum-subtracted rest frame $0.4-1.2\1{mm}$ spectrum of high redshift submillimeter galaxies, constructed from 78 SPT DSFGs and shown at $500\1{km/s}$ resolution.  Lines given with bold font represent $>3\sigma$ detections using the stacking procedure described in Sec.~\ref{sec:stacked_spec}, while the other lines indicate potentially detectable molecular lines ($1.5\sigma < \mathrm{S/N} < 3\sigma$).  The running $\pm 1 \sigma$ noise level is shown in blue.
	The middle panel shows the running signal to noise ratio of the top panel and the bottom panel shows the number of sources which contribute at each frequency. }
	\label{fig:stacked_spec}
\end{figure*}

We show the average $250-770\1{GHz}$ spectrum of the SPT DSFG sample in Fig.~\ref{fig:stacked_spec}, with the detected and potentially detected lines indicated.  Due to the majority of sources lying in the redshift range $2.5 < z < 6.0$, the stacked spectrum is the most sensitive to the rest frame frequency range $300-800\1{GHz}$.  In order to properly assess the significance of these lines, we consider two factors: the statistical uncertainty of the individual ALMA spectra, and the intrinsic scatter of the underlying source population.  Like the flux density, the statistical uncertainty of the individual ALMA spectra is scaled according to each source's dust mass in order to preserve the signal to noise ratio.  

As noted in \citet{spilker14}, the stacked spectrum is not ideal for the detection of individual molecular lines since the regular channel grid will likely split the line flux of a given line over multiple channels.  In order to assess the flux density of these individual lines, spectra were constructed and centered on the line of interest, with $600\1{km/s}$ channels.  These spectra were then stacked according to the procedure detailed in Sec.~\ref{sec:stackmethods}, and a summary of detected lines and upper limits is given in Tab.~\ref{tab:lines}.  In the case where two lines are blended, as in the cyanide radical CN(J=4-3) and the nearby HNC(J=5-4) line, we use the average of the line ratios obtained in \citet{guelin07, bethermin18, canameras21} to determine the relative contributions of the individual lines.  

For many species, individual transitions fall below the $3\sigma$ detection threshold.  However, by stacking all available transitions of a molecular species, we can constrain the total luminosity emitted by the molecule in the covered rest frequency range.  In Tab.~\ref{tab:stackedlines}, we give the total luminosity in each species for all transitions that happen to fall within our frequency range.  The bright molecular lines $^{12}$CO, H$_2$O and $[$CI$]$ are all among the most luminous species observed in the ALMA $3\1{mm}$ bandpass.  At first glance, it is surprising that H$_2$O$^+$ is more luminous than H$_2$O in this bandpass, given that in both local and high redshift examples (e.g.~\citealp{yang13, yang16, rangwala11}) the H$_2$O lines are typically $\sim 50\%$ more luminous than H$_2$O$^+$ lines.  However, a greater number of H$_2$O$^+$ transitions are accessible from the ALMA $3\1{mm}$ bandwidth compared to those of H$_2$O, and the most luminous H$_2$O transitions are well outside of our window (e.g. H$_2$O(2$_{0,2}$-1$_{1,1}$) at $988\1{GHz}$).  These considerations, along with the large uncertainties on total luminosity for both molecular species, help explain the higher total luminosity of H$_2$O$^+$.

\begin{deluxetable}{lcccc} 
\tabletypesize{\small}
\tablewidth{0pc}
\tablecolumns{4}
\tablehead{
\colhead{Species} & \colhead{Transitions} & \colhead{Observations} & \colhead{$\sum$L}  \\ 
 & & & ($10^8$ L$_\odot$) \\ 
}
\startdata 
$^{12}$CO   &   4   & 	102     & 	39.61 $\pm$ 1.99     \\
H$_2$O$^+$ 	&	8 	&   124 	&   6.11 $\pm$	1.99     \\
H$_2$O    	&	5 	&   116 	&   5.39 $\pm$ 	1.02     \\
$[$CI$]$    &	1 	&   38 	    &   1.98 $\pm$ 	0.06     \\
HCN   		&   6 	&   124 	&   1.57 $\pm$ 	0.84     \\
$^{13}$CO  	&	4 	&   101 	&   1.16 $\pm$ 	0.31     \\
HCO$^+$  	&	6 	&   127 	&   1.08 $\pm$ 	0.91     \\
N$_2$H$^+$  &	6 	&   121 	&   1.07 $\pm$ 	1.12     \\
CH   		&   2 	&   74 	    &   0.90 $\pm$ 	0.10     \\
HNC   		&   6 	&   125 	&   0.47 $\pm$ 	0.91     \\
OH   		&   2 	&   66 	    &   0.11 $\pm$ 	0.06     \\
H$\alpha$ RRLs  &   8   &   149 & 	-0.12 $\pm$ 0.31     \\
CN    		&   4 	&   99 	    &   -0.59 $\pm$ 0.43     \\
C$^{18}$O  	&	5 	&   101 	&   -1.16 $\pm$ 1.22     \\
SiO    		&   12 	&   254 	&   -1.81 $\pm$ 1.17     \\
CS     		&   9 	&   195 	&   -1.85 $\pm$ 1.05     \\
\enddata 
\caption{Ranked order of most luminous molecular cooling lines for the ALMA $3\1{mm}$ bandpass for the redshift range $1.9 < $z$<6.9$.  These lines are ranked according to the total luminosity ($\Sigma$L) summed across all transitions in our observed frequency range.\label{tab:stackedlines} }
\end{deluxetable}

\section{Discussion} \label{sec:discussion}			%

\subsection{$^{12}$CO SLED} \label{sec:12cosled}


\begin{figure*}[!th]
	\centering
	\includegraphics[width=\textwidth]{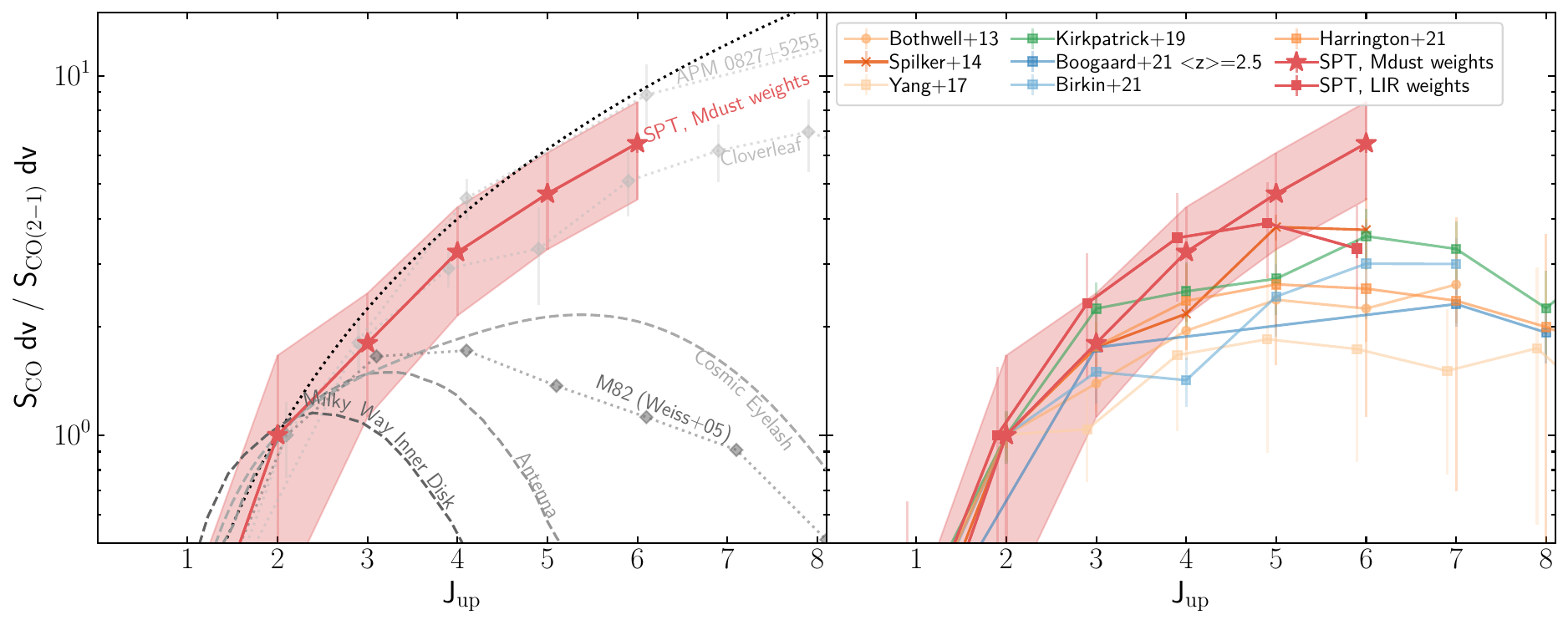}
	\caption{\footnotesize $^{12}$CO SLEDs of the SPT DSFG sample, normalized to the $J=2-1$ transition, shown along with individual sources (top panel) and composite SLEDs from literature (bottom panel).  \textit{Bottom panel:} SLEDs from surveys that were largely unlensed are shown in blue, while lensed samples are in orange.  The \citet{kirkpatrick19} contains a combination of both unlensed and lensed sources, and is shown in green.  The marker type denotes the normalization method, where squares, x's and stars correspond to normalization by L$_{\mathrm{IR}}$, $1.4\1{mm}$ flux and dust mass respectively.  While \citet{bothwell13} is also normalized by L$_{\mathrm{IR}}$, the median rather than the average was used and it is denoted by a circle. }
	\label{fig:sled}
\end{figure*}

Using the $^{12}$CO fluxes taken from the composite spectrum, we construct a spectral line energy distribution (SLED) shown in Fig.~\ref{fig:sled}, in comparison with both well-sampled SLEDs of individual objects as well as other statistical SLEDs.  The Milky Way inner disk~\citep{fixsen99} and Antennae Galaxies~\citep{zhu03} represent objects in our local Universe.  However, the high redshift, gravitationally lensed objects H1413+117 (the ``Cloverleaf" quasar at z=2.56; \citealp{bradford09}), SMM J2135-0102 (the ``Cosmic Eyelash" DSFG at z=2.32; \citealp{danielson11}) and APM 0827+5255 (a quasar at z=3.91; \citealp{weiss07}) are more analogous to the DSFGs in the SPT sample.  While the full SPT sample normalized by L$_{\mathrm{IR}}$ (see bottom panel of Fig.~\ref{fig:sled}) shows an apparent flattening around J=5, analogous to local starburst galaxies M82 or NGC 253~\citep{panuzzo10, bradford04}, 
normalization by IR luminosity will affect the overall shape of the $^{12}$CO SLED, since the excitation is dependent on IR luminosity.  
The intrinsic scatter of the full SPT sample is denoted by the shaded region around the dust mass weighting and suggests that while there is a large variation in excitation, the excitation is similar to QSOs.  

Rather than observe a single object for as many CO transitions as possible, statistical samples can be built through observing a large sample of sources with only a few or even single CO transitions.  There are a wide variety of ways to combine these data (see Sec.~\ref{sec:stackmethods} for a full discussion) in order to create a composite SLED.  The first composite SLEDs were compiled in \citet{bothwell13} and \citet{spilker14} and have since been done for a variety of samples, including \citet{yang17}, \citet{valentino20}, \citet{boogaard21} and \citet{birkin21}.  
 
However, because the SLEDs shown are constructed with different normalizations, it is difficult to compare the differences directly.  
While \citet{bothwell13} and \citet{spilker14} both utilize populations of DSFGs, they normalize by far-IR luminosity and $1.4\1{mm}$ flux, respectively.  As discussed in Sec.~\ref{sec:stackmethods}, the normalization quantity can make a $25\%-65\%$ difference in the line ratio.  While \citet{yang17}, \citet{boogaard21} and \citet{birkin21} all normalize by L$_{\mathrm{IR}}$, \citet{yang17} is largely composed of lensed DSFGs, while \citet{boogaard21} and \citet{birkin21} are primarily comprised of unlensed DSFGs.  The sample of sources presented in \citet{kirkpatrick19} contains between $58-75\%$ strongly lensed sources ($\mu > 5$; percentage dependent on J transition), which is similar to the $\sim 70\%$ of strongly lensed sources ($\mu > 8$; \citealp{spilker16}) found in the SPT sample.  In spite of the choice of normalization and large intrinsic scatter of the SPT sample, the average of the SPT DSFGs does appear more energetic than many other DSFG SLEDs in the literature.  

\begin{figure}[!hbt]
	\centering
	\includegraphics[width=\columnwidth]{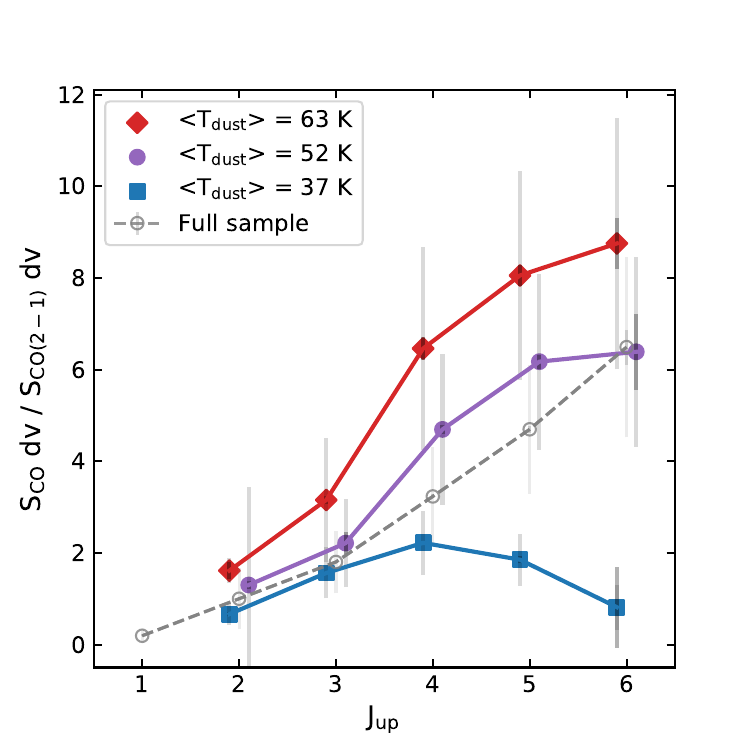}
	\caption{The SPT DSFG sample split into three equal sub-samples according to dust temperature, all normalized to the J$_{\mathrm{up}}=2$ flux density of the full sample.  The higher dust temperature systems contain more excited gas, while the lower dust temperature systems exhibit sub-thermal behavior and show an apparent flattening near J$_{\mathrm{up}}=5$, like the local star-bursting systems M82 and NGC253~\citep{panuzzo10, bradford03}.  }
	\label{fig:cosledtdust}
\end{figure}

The excitation of the bright $^{12}$CO emission lines has been shown to correlate well with dust temperature~\citep{rosenberg15}.  In order to test the effects of dust temperature on the shape of the SPT composite SLED, the SPT sample was split into three equal sub-samples according to dust temperatures from~\citet{reuter20}.  The same stacking procedure was used for each sub-sample, and the resulting SLEDs are shown in Fig.~\ref{fig:cosledtdust} in comparison with the full sample.  Because DSFGs are some of the most extreme star-forming galaxies, the hottest sources in the SPT sample exhibit much more excitation in the higher-J transitions than the lower dust temperature sources, which show an apparent flattening around J=5, like the local starburst galaxies.  

One important caveat is that the lower-J lines tend to be populated by lower redshift sources at lower dust temperature, while the higher-J lines tend to have higher redshift and dust temperature sources.  While \citet{reuter20} found a slight correlation of dust temperature with redshift, the intrinsic scatter was sufficiently large that this effect was not statistically significant.  However, the difference between the high and low dust temperature SLEDs can also be observed at intermediate dust temperature (shown in Fig.~\ref{fig:cosledtdust} as purple circles) and through J=3-5, where there is a more equal distribution of sources between the different temperature bins.  

\subsection{$^{13}$CO and C$^{18}$O } \label{sec:13coandc18o}
Though $^{12}$CO and its isotopologues, $^{13}$CO and C$^{18}$O, are all thought to originate from the same volume~(e.g. \citealp{henkel10, zhang18b}), $^{12}$CO is optically thick even at moderate densities ($\mathrm{n_{H_2}} \lesssim 10^3 \1{cm^{-3}}$), due to its high relative abundance and low dipole moment.  However, the rare $^{13}$CO and C$^{18}$O isotoplogues are typically optically thin due to their low abundance and are therefore capable of probing the total molecular column density of cold gas regions~\citep{carilli13}.  While both $^{12}$C and $^{13}$C are formed in and ejected from stars, $^{12}$C is generally produced in high-mass stars at rapid timescales.  In contrast, $^{13}$C is a ``secondary" species, since it is produced in longer-lived, low to intermediate mass stars and its creation involves enriching $^{12}$C seed nuclei~\citep{wilson94}.  Since the creation of $^{13}$C requires the ISM to be enriched by metals from previous generations of stars, the relative abundances of $^{12}$C and $^{13}$C are indications of the star formation history of a system.  The emission strengths of the $^{12}$CO and $^{13}$CO lines are related to the abundances of their respective carbon isotopes, given a known optical depth.  $^{13}$CO is expected to be optically thin under most conditions, except the most dense, star-forming cores.  Less is known about the formation of $^{18}$O, though it is likely formed through the CNO cycle in massive stars and is enhanced in galaxies with recent massive star formation~\citep{henkel93}.  Due to its lower abundance and optical depth compared to $^{13}$CO, C$^{18}$O can be used to probe different gas column densities, provided the elemental abundances remain constant.  

In the SPT DSFG composite spectrum, we detect $^{13}$CO at $>3\sigma$ for the J$_{\mathrm{up}} = 3-5$ transitions and set limits for C$^{18}$O.  While the exact luminosities and significances can be found in Tab.~\ref{tab:lines}, we find that the line ratio between the $J=3-2$ and $J=4-3$ lines increases, with $L'_{^{12}\mathrm{CO}}/L'_{^{13}\mathrm{CO}} \sim 20$ and  $L'_{^{12}\mathrm{CO}}/L'_{^{13}\mathrm{CO}} \sim 30$ for $J=3-2$ and $J=4-3$, respectively.  This is likely due to a lower excitation in $^{13}$CO due to less line trapping, which is related to lower optical depth.  Past $J=4-3$ however, the $L'_{^{12}\mathrm{CO}}/L'_{^{13}\mathrm{CO}}$ ratio seems to remain constant within errors.  However our measurements are only sensitive enough to constrain until the $J=5-4$ transition.  

The $^{12}$C/$^{13}$C abundance ratios have been studied in the Milky Way using a variety of different molecules and have generally been found to steadily increase from $\sim 20$ near the galactic center to $\sim 70$ in our solar system (e.g.~\citealp{langer90,aalto91, aalto95, wilson95}), likely due to differences in optical depth as one moves from the galactic center~\citep{milam05}.  In regions of high star formation, such as giant molecular clouds, the $^{12}$C/$^{13}$C ratio is reported to be between $\sim3-5$, with more $^{13}$C present in the clouds' central regions~\citep{solomon79, polk88}.  A recent survey of the J$_{\mathrm{up}}=1$ line for 147 nearby main-sequence galaxies revealed an integrated intensity ratio of 10.9 with a standard deviation of 7.0~\citep{morokumnanatsui20}, while local star-forming and starbursting galaxies exhibit higher ratios of $16.1 \pm 2.5$~\citep{mendezhernandez20}.  Taken together, these surveys imply that galaxies with increasing levels of star formation exhibit increasingly suppressed levels of $^{13}$C, but there is a large intrinsic scatter in the sample.  Indeed, some highly star-forming systems have been found to have brightness temperature $^{12}$CO/$^{13}$CO ratios of $\sim 40$ or even as high as $\geq 60$~\citep{young21, sliwa17}.  Additional studies have shown that the $^{12}$C/$^{13}$C ratio is sensitive to environmental conditions such as the density of nearby galaxies~\citep{alatalo15c} and star formation rate surface density~\citep{davis14}.  

At high redshift ($z>2$), samples with $^{13}$CO and C$^{18}$O are limited to observations of the bright CO transitions in gas- and metal-rich galaxies amplified with gravitational lensing~\citep{henkel10, danielson13, spilker14, bethermin18, zhang18b}.  Typical values of the $^{12}$C/$^{13}$C ratio range from $\sim 20-40$ for the most commonly observed $J=4-3$ transition and the $J=4-3$ ratio obtained from the composite spectrum, $L'_{^{12}\mathrm{CO}}/L'_{^{13}\mathrm{CO}} \sim 30$, is consistent with that picture.  

Because C$^{18}$O is also associated with high star formation rate regions, $^{13}$CO/C$^{18}$O line ratios have been shown to differ in galaxies with differing star formation rates.  In the local Universe, typical $^{13}$CO/C$^{18}$O ratios range from $\sim 4-8$ (e.g.~\citealp{henkel93}).  However, in local ULIRGs and starbursts, ratios typically range from $\sim 1-2$~\citep{greve09, sliwa17, mendezhernandez20}.  
While we do not formally detect C$^{18}$O, the limits on the  $^{13}$CO/C$^{18}$O ratio range from $\sim 2-3$ with an average of ratio $3.2$ across all transitions.  While these values are smaller than the typical local $^{13}$CO/C$^{18}$O ratios, they are slightly higher than those found in high redshift ULIRGs and starbursts (e.g.~\citealp{danielson13}).  
However, one cannot directly use the observed abundance ratios alone to draw conclusions about the underlying abundances.  Detailed radiative transfer calculations are needed to further interpret these results.  

\subsection{Dense Gas Tracers} \label{sec:densegas}
Though $^{12}$CO traces the total molecular gas, observations of nearby galaxies have found a close link between the dense gas and star formation activity~(e.g.~\citealp{gao04a, jimenez-donaire19}).  Molecules with higher critical densities and excitation energies such as CN, HCN, HNC and HCO$^+$ have been used as probes of the warm, dense molecular gas where stars form.  In the SPT DSFG composite spectrum, we detect the $J=4-3$ transitions of HCN, HNC, and HCO$^+$, along with the $J=5-4$ transition of HNC and the $J=3-2$ transition of CN.  While the $J=4-3$ transition of the CN radical was detected in \citet{spilker14}, it was only marginally detected with a significance of $2.9\sigma$ in this work.  Along with the differences in stacking method detailed in Sec.~\ref{sec:stackmethods}, this line was also blended with the $J=5-4$ transition of HCN (the de-blending process is discussed in Sec.~\ref{sec:stacked_spec}), which also contributed to the decreased significance.  

\begin{figure}[h]
	\centering
	\includegraphics[width=\columnwidth]{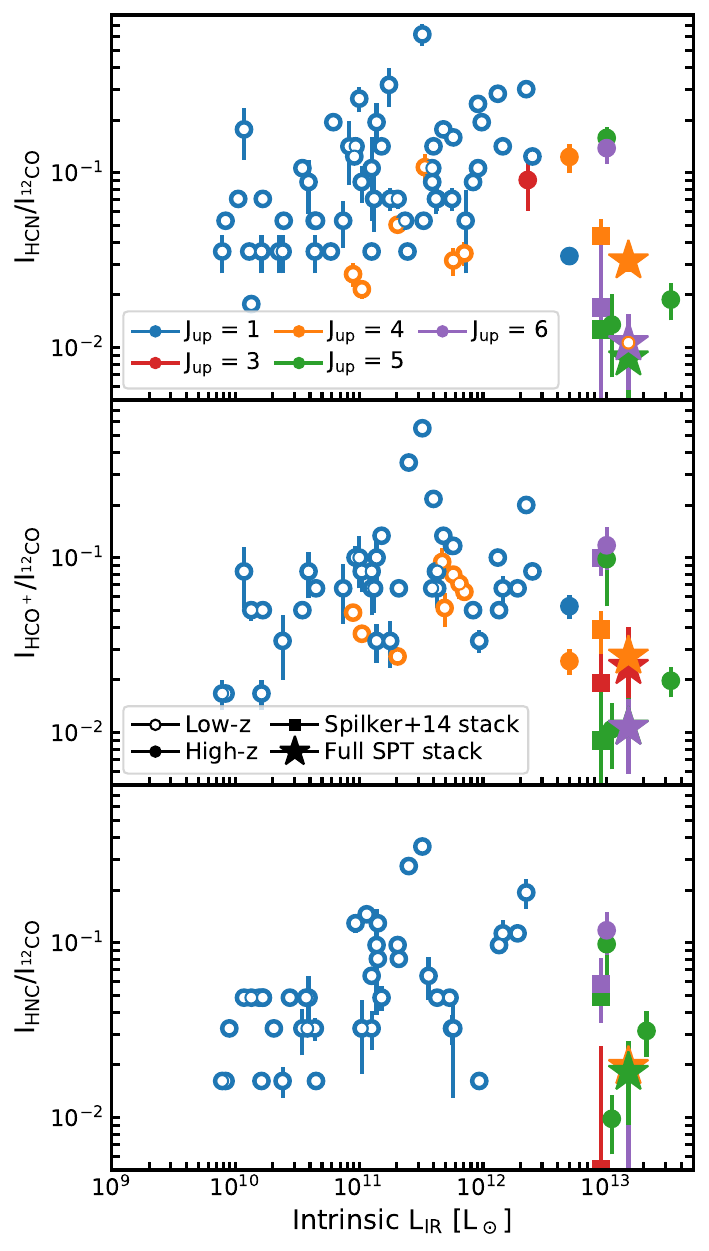}
	\caption{Line ratios between the dense gas tracers, HCN, HCO$^+$ and HNC, and $^{12}$CO as a function of the intrinsic L$_{\mathrm{IR}}$.  Local LIRG and ULIRG samples are represented by the hollow symbols, while the high redshift samples are solid.  Ratios obtained from single sources are represented by circles, while ratios from stacked spectra are generally represented by the squares and stacked spectra presented in this work is denoted by the star. }
	\label{fig:densegasratio_v_lir}
\end{figure}

The flux densities of the dense gas tracers HCN, HNC and HCO$^+$ are shown in Fig.~\ref{fig:densegasratio_v_lir} compared with the $^{12}$CO flux density as a function of the intrinsic IR luminosity from \citet{reuter20}.  The apparent IR luminosity was determined by fitting the photometry published in \citet{reuter20} according to a modified blackbody with an additional power law component.  The intrinsic IR luminosity was then obtained by using either the magnification values from detailed lens models~\citep{spilker16} or by using the median magnification factor of $\mu_{870\1{\mu m}} = 5.5$.  Comparison samples include nearby LIRGs and ULIRGs from a variety of surveys~\citep{baan08, zhang14} and a few measurements from high redshift lensed objects~\citep{guelin07, riechers07, riechers10, danielson13, bethermin18, canameras21} as well as average values obtained from stacking~\citep{spilker14}.  The rotational ground levels of the dense gas molecules compared to CO are proxies for the fractional value of dense gas to total mass in the system, while the IR luminosity is proportional to the star formation rate~(e.g. \citealp{murphy11}).  Though some excited rotational ground levels are also shown in Fig.~\ref{fig:densegasratio_v_lir}, it is difficult to compare because of the different excitation requirements for $^{12}$CO and the dense gas tracers shown.  Recent work~\citep{rybak22} has found that (U)LIRGs at high redshift have a lower HCN/CO than the local Universe for the J$_{\mathrm{up}}=1$ transition.  Though our results seem to also show lower HCN/CO ratios at high redshift, the SLED of HCN is a subject of ongoing study in the local Universe (e.g.~\citealp{saito18}) and has not been explored at high redshift.  

In the local Universe, it has been shown that the ground state HCN emission strongly correlates with IR luminosity over many orders of magnitude~\citep{gao04a, usero15, bigiel16}, but it is unknown if this correlation continues at high redshift.  At Cosmic Noon ($2<z<4$), the star formation rate was $\sim 10\times$ higher than its present day level and is generally not well represented by the modes of star formation seen in nearby galaxies~\citep{madau14, genzel11}, except perhaps in rare local ULIRGs.  It is unclear if the prodigious star formation rates observed in high redshift galaxies are due to increased gas accretion, or if an increase in the efficiency of the star formation process is required.  One possibility is that at high redshift, the star formation rate still scales linearly with IR luminosity, but the dense gas fraction is enhanced, driven by some combination of gravitational instabilities from mergers and gas accretion, and extreme stellar feedback from supernovae or radiation pressure~\citep{bournaud10}.  However, recent work~\citep{rybak22} indicates that the dense gas fraction is suppressed rather than enhanced.  Another possibility is that the efficiency with which dense gas forms stars depends on the local conditions of the galaxy~\citep{usero15, bigiel16}.  More observations, especially of HCN, are necessary to more fully understand the relationship between dense gas and star formation at high redshift.  

The ratios between dense gas tracers can also constrain the physical conditions of the gas.  For instance, the HNC/HCN and HCO$^+$/HCN ratios have been suggested to provide a discriminator between photo-dissociation regions (PDRs) and X-ray-dominated regions (XDRs) or more complex dynamics~\citep{meijerink07, aalto07}, though X-ray driven chemistry can change abundance ratios (e.g.~\citealp{juneau09}).  For the $J=4-3$ transition, we find a HNC/HCN ratio of $\sim 0.6$ and a HCO$^+$/HCN ratio of $\sim 0.9$, which are lower than previously obtained values in \citet{spilker14} and for individual sources in \citet{bethermin18}.  However, the obtained ratios are in agreement with values for local ULIRGs (HNC/HCN $\sim 0.5$ and HCO$^+$/HCN $\sim 0.8$ for $J=1-0$ detections; \citealp{jimenez-donaire19}).  Averaging over all transitions gives ratios of $\sim 0.9$ and $\sim 1.0$ for HNC/HCN and HCO$^+$/HCN respectively, in agreement with both previously obtained SPT values and measurements from other high redshift lensed galaxies~\citep{canameras21}.  For the HNC/HCN ratio, measurements are in the typical range of PDRs (HNC/HCN $\simeq$ 0.1-1.0; \citealp{hirota98, hacar20}), and the differences between the overall HNC/HCN ratio and the individual transitions suggest that the HNC/HCN ratio has some dependence on the rotational quantum number.  In contrast, since the $J=4-3$ and averaged measurements for the HCO$^+$/HCN ratio agree within errors and are consistent with other measurements at high redshift, the HCO$^+$/HCN ratio does not appear to exhibit any frequency dependence.  

\begin{figure*}[htb]
	\centering
	\includegraphics[width=\textwidth]{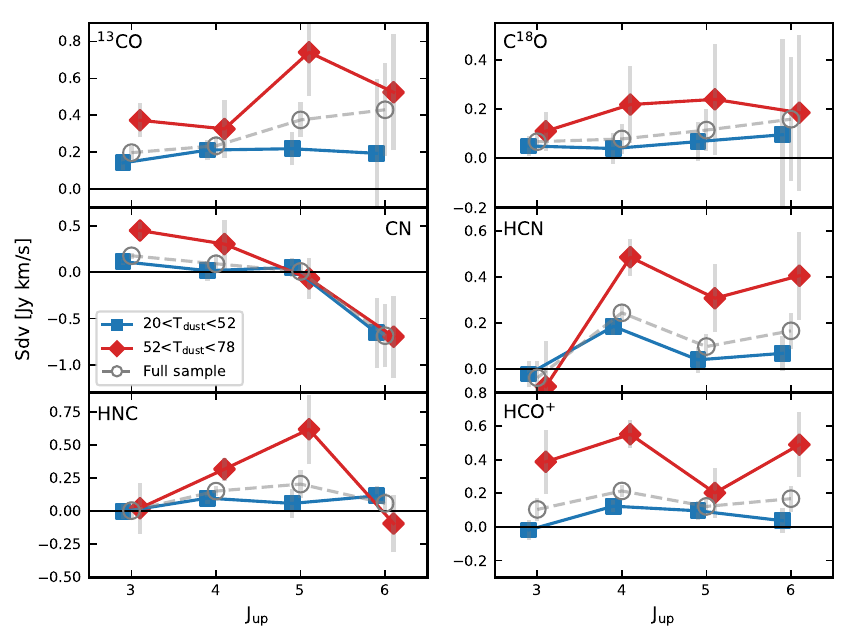}
	\caption{Spectral line distributions for $^{13}$CO, C$^{18}$O, CN, HCN, HNC and HCO$^+$ from the SPT DSFG composite spectrum.  The sources are ordered by dust temperature and split into hot (red diamonds; $<$$\mathrm{T_{dust}}$$>$$= 42\1{K}$) and cold (blue squares; $<$$\mathrm{T_{dust}}$$>$$= 61 \1{K}$) populations.  The full sample is represented by the hollow circles and dashed line. 
	\label{fig:faintlinesled}}
\end{figure*}

As in Sec.~\ref{sec:12cosled}, we again order the SPT DSFG sample by the dust temperatures from \citet{reuter20} and split the sample into two bins according to temperature, as shown in Fig.~\ref{fig:faintlinesled}.  While the majority of the transitions of the optically thin $^{13}$CO and C$^{18}$O exhibit differences between the so-called ``hot" and ``cold" DSFG populations, these populations are consistent within errors to the full sample.  The dense gas tracers HCN, HNC and HCO$^+$ for hotter DSFGs are potentially more energetic in some transitions than their cooler counterparts.  Though more observations are needed to increase the statistical robustness of the sample, this could indicate the presence of more excited dense gas or a more massive dense gas reservoir in hotter DSFGs.  Additionally, the ``hot" and ``cold" DSFG populations show no difference for the cyanide radical, CN, and all three populations show indications of absorption in the J$_{\mathrm{up}}=6$ transition.  It should be emphasized that these results are speculative since many of these transitions have not been detected in this work and are also dependent on the normalization (see Sec.~\ref{sec:stackmethods} for details).  However, as orders of magnitude more sources are expected to be discovered with SPT-3G, these results could highlight interesting future areas of study.  

\subsection{Water} \label{sec:water}
Like the molecules discussed in Sec.~\ref{sec:densegas}, H$_2$O emission arises from warm gas in dense molecular filaments and structures.  In the SPT-DSFGs, the H$_2$O(2$_{1,1}$-2$_{0,2}$) and H$_2$O(4$_{1,4}$-3$_{2,1}$) emission lines ($\nu_{\mathrm{rest}} = 752\1{GHz}$, $E_{\mathrm{up}}=137\1{K}$ and $\nu_{\mathrm{rest}} = 380\1{GHz}$, $E_{\mathrm{up}}=324\1{K}$, respectively) are detected, along with the H$_2$O(1$_{1,0}$-1$_{0,1}$) absorption line ($\nu_{\mathrm{rest}} = 557\1{GHz}$, $E_{\mathrm{up}}=61\1{K}$; see Tab.~\ref{tab:lines} for details).  The H$_2$O(2$_{1,1}$-2$_{0,2}$) emission line is excited through the absorption and pumping of the $101\1{\mu m}$ transition line and traces the infrared field.  Studies~\citep{omont13, yang13, gonzalezalfonso14b} show that this line is correlated with IR luminosity over several orders of magnitude in both local and high-z LIRGs and could also be used to trace star formation (e.g. \citealp{jarugula19}).  Because the H$_2$O(4$_{1,4}$-3$_{2,1}$) emission line is a higher energy transition, it has been detected in very dense regions, such as in the circumnuclear disk around a quasar (e.g. \citealp{stacey20}).  Since this line was found to be spatially coincident with the CO(11-10) emission line, it is likely that H$_2$O(4$_{1,4}$-3$_{2,1}$) is found in highly excited regions, like high-J CO lines.  Radiative transfer models (e.g.~\citealp{liu17}) also suggest that the  H$_2$O(4$_{1,4}$-3$_{2,1}$) requires extreme (T$_{\mathrm{dust}} \sim 100-200$) conditions in order to be excited.  Given the highly excited CO line fluxes shown in Fig.~\ref{fig:sled}, at least a subset of the SPT DSFGs exhibit such conditions.  

Though the H$_2$O(4$_{2,3}$-3$_{3,0}$) emission line (at $448\1{GHz}$) was detected in a strongly lensed DSFG at $z=3.6$ (G09v1.97; \citealp{yang20}), it has not been observed in \citet{spilker14} or this work.  Given the detection of the H$_2$O(2$_{1,1}$-2$_{0,2}$) emission line (at $752\1{GHz}$) in both \citet{yang20} and in the SPT DSFG sample, and the inferred line ratios from \citet{yang20}, the H$_2$O(4$_{2,3}$-3$_{3,0}$) emission line (at $448\1{GHz}$) should have been detected in this work \textit{if} G09v1.97 was representative of the larger population.  Additionally, the SPT DSFG sample is on average more intrinsically luminous than G09v1.97 (SPT median intrinsic luminosity is $\langle \mathrm{L}_{\mathrm{IR}}\rangle = 1.5 \times 10^{13}\1{L_\odot}$ compared to L$_{\mathrm{IR}} = 5 \times 10^{12}\1{L_\odot}$), which could change the excitation levels in the water transitions.  

\begin{figure}[h]
	\centering
	\includegraphics[width=\columnwidth]{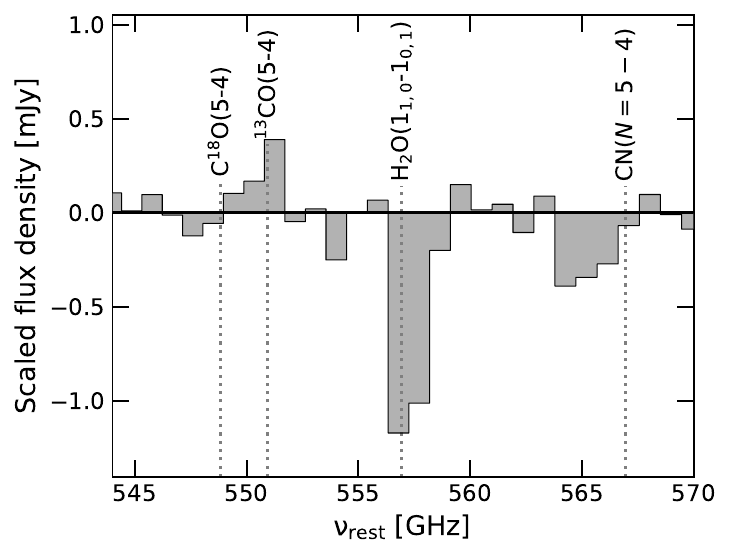}
	\caption{The H$_2$O(1$_{1,0}$-1$_{0,1}$) absorption line from the composite spectrum, given in flux density and normalized by the dust masses of individual contributing sources.  This transition is observed in the composite spectrum with an equivalent width of $199 \pm 24\1{km/s}$.}
	\label{fig:h2oabsorb}
\end{figure}

A deep absorption feature due to the H$_2$O(1$_{1,0}$-1$_{0,1}$) ground state transition was also observed in the SPT DSFG stack.  Though we perform a continuum subtraction in order to obtain the stacked spectrum, we measure the continua of the input spectra beforehand and create a stacked continuum in order to determine the equivalent width.  We measure an equivalent width of $199 \pm 24\1{km/s}$ for the H$_2$O(1$_{1,0}$-1$_{0,1}$) transition measured in our stacked spectrum.  This is the only absorption line securely detected in our stack, and indicates the presence of some low excitation gas backlit against the continuum, along the line of sight.  This absorption line has also been seen in \citet{riechers22} for a DSFG at z=$6.34$, where the absorption feature was detected at the $\sim 2\sigma$ level, relative to the CMB, indicating the presence of cold water vapor.  However, more study would be needed to determine if this effect is also occurring in the individual DSFGs present in our stacked spectrum.   

\subsection{Carbon hydrides} \label{sec:CH}
The carbon hydride CH has been observed locally in both dense photodissociation regions (e.g.~\citealp{habart10}) and diffuse molecular clouds (e.g.~\citealp{chastain10}).  
Because CH exhibits constant abundance ratios relative to molecular hydrogen (e.g.~\citealp{liszt02}), it is considered to be a reliable tracer of molecular hydrogen.  
Additionally, because neutral hydrides like CH are most abundant in gas with a large molecular fraction~\citep{gerin16}, measurements of this molecule can be combined with other diagnostic lines to determine the molecular fraction.  
CH was first detected at high redshift in \citet{spilker14} by combining all transitions in the $3\1{mm}$ window.  It has again been detected in the stacked spectrum presented in this work and also at $532\1{GHz}$ and $536\1{GHz}$.  
It has been suggested that the CH doublet is sensitive to both the fine structure constant and electron mass ratio~\citep{denijs12} and measurements of CH at $532$ and $536\1{GHz}$ could provide a test of these fundamental constants with time.  However, precise calibration with another line (e.g. H$_2$O(1$_{1,0}$-1$_{0,1}$) at $557\1{GHz}$) would be needed to study these effects.  

\section{Summary and Conclusions} \label{sec:conclusions}		%
We have presented the average rest-frame millimeter spectrum of the complete SPT DSFG sample.  By stacking the $3\1{mm}$ spectra obtained with ALMA Band 3 for a total of 78 objects from $z=1.9-6.9$, we are able to probe faint ISM diagnostic lines and study the characteristics of high redshift DSFGs.  Our conclusions are as follows:
\begin{enumerate}
\item Using the composite SLED, we are able to detect not only the very bright $^{12}$CO and [CI] emission lines, but also multiple fainter molecular transitions from $^{13}$CO, C$^{18}$O, HCN, HNC, HCO$^+$, CN, H$_2$O, CH, and OH.  Both the obtained detections and the limits set for other molecules are invaluable diagnostics of the ISM and will serve as guides to plan future ALMA observations.  
\item The bright $^{12}$CO transitions from J$_{\mathrm{up}} = 3-6$ were combined with low-J ATCA observations in order to produce a SLED similar to, but more excited than, many comparable SMG samples.  Because dust temperature has been shown to correlate with the CO SLED shape (e.g.~\citealp{rosenberg15}) and the SPT sample tends to have higher dust temperatures than comparable samples~\citep{reuter20} it is not surprising that the CO SLED is more excited than others.  However, we highlight the importance of using comparable weights when comparing other composite spectra.  To further investigate this temperature dependence, we split the sample according to dust temperature, revealing a stark difference between the lower temperature systems, which exhibit sub-thermal behavior and show an apparent flattening near J$_{\mathrm{up}} = 5$ and higher temperature systems near local thermal equilibrium.  
\item The J$_{\mathrm{up}} = 3-5$ transitions of $^{13}$CO were detected, and limits were placed on J$_{\mathrm{up}} = 6$, along with all accessible transitions of C$^{18}$O.  However, constraints set on the $^{12}$CO/$^{13}$CO and $^{13}$CO/C$^{18}$O ratios are largely consistent with previously obtained ratios at high redshift and indicate that the DSFGs at high redshift exhibit suppressed levels of $^{13}$C compared to local ULIRGs, though more observations would be needed to conclude anything about $^{18}$O.  
\item Spectral line distributions were created using the additional detections of $^{13}$CO, C$^{18}$O, HCN, HNC, HCO$^+$ and CN.  Splitting the spectra according to dust temperature revealed that hotter systems appear to either contain warmer and more excited dense gas tracer (HCN, HNC and HCO$^+$) transitions or simply more dense gas.  However, there is no difference between hot and cold systems for the optically thin $^{13}$CO and C$^{18}$O molecules.  
\item We also robustly detect the  H$_2$O(2$_{1,1}$-2$_{0,2}$) and H$_2$O(4$_{1,4}$-3$_{2,1}$) emission lines, as well as the H$_2$O(1$_{1,0}$-2$_{0,1}$) absorption line.  While the emission lines trace the densest regions of DSFGs and can potentially be used as indicators of star formation, the presence of the absorption line detected at $557\1{GHz}$ indicates that there could be reservoirs of lower excitation gas, backlit against the continuum along the line of sight.  
\item Though this is the largest composite spectrum of DSFGs to date, orders of magnitude more sources are expected to be discovered with SPT-3G~\citep{benson14} and CMB-S4~\citep{abazajian19}.  Spectroscopic follow-up of these sources will be able to uncover the earliest stages of the dust-obscured Universe. 
\end{enumerate}

\vspace{5mm}
\facilities{ALMA, ATCA, SPT}

\software{
    \texttt{astropy}~\citep{astropy:2013, astropy:2018, astropy:2022},
    \texttt{CASA}~\citep{mcmullin07, petry12}
    }
          
\acknowledgments
The SPT is supported by the NSF through grant OPP-1852617.
D.P.M. and J.D.V.acknowledge support from the US NSF under grants AST-1715213 and AST-1716127.
J.D.V. acknowledges support from an A.~P.~Sloan Foundation Fellowship. 
M.A. acknowledges support from FONDECYT grant 1211951, CONICYT + PCI + INSTITUTO MAX PLANCK DE ASTRONOMIA MPG190030, CONICYT + PCI + REDES 190194, and ANID BASAL project FB210003.  
M.A.A. and K.A.P are supported by the Center for AstroPhysical Surveys at the National Center for Supercomputing Applications as Illinois Survey Science Graduate Fellows.  
The National Radio Astronomy Observatory is a facility of the National Science Foundation operated under cooperative agreement by Associated Universities, Inc.
This paper makes use of the following ALMA data: 
ADS/JAO.ALMA\#2011.0.00957.S, 
ADS/JAO.ALMA\#2012.1.00844.S, 
ADS/JAO.ALMA\#2015.1.00504.S, 
ADS/JAO.ALMA\#2016.1.00672.S and 
ADS/JAO.ALMA\#2019.1.00486.S. ALMA is a partnership of ESO (representing its member states), NSF (USA) and NINS (Japan), together with NRC (Canada), MOST and ASIAA (Taiwan), and KASI (Republic of Korea), in cooperation with the Republic of Chile. The Joint ALMA Observatory is operated by ESO, AUI/NRAO and NAOJ.
The Cosmic Dawn Center (DAWN) is funded by the Danish National Research Foundation under grant No. 140.

\appendix						    %

\section{Stacked spectrum weights}
\label{ap:weights}
\renewcommand{\theHtable}{A.\arabic{table}}
\renewcommand{\theHfigure}{A.\arabic{figure}}
\setcounter{table}{0}
\setcounter{figure}{0}
\renewcommand\thetable{\thesection.\arabic{table}}  
\renewcommand\thefigure{\thesection.\arabic{figure}} 

The weights used in the stacked spectrum are given in Tab.~\ref{ap:weights}.  The spectra were first re-scaled to a common rest frame (described in Sec.~\ref{sec:stackmethods} and then normalized by the apparent dust mass of the source, as a proxy for size.  The apparent dust mass was calculated with the same photometry and method described in \citet{reuter20}, except was calculated using $\kappa_{850\1{\mu m}} /$ $\1{m^2}$ $\1{kg^{-1}} = 0.038$~\citep{draine03} instead of the value given in \citet{reuter20}.  However, because the weights are scaled according to the average apparent dust mass of the sample, ($\langle \mathrm{M}_{\mathrm{dust}}\rangle = 8.7 \times 10^9 \, \mathrm{M}_\odot$), the value of $\kappa$ does not contribute to the relative weight of an individual source in the stack.  We also give the average noise across a $62.5\1{MHz}$ bandwidth in Tab.~\ref{ap:weights} for the $3\1{mm}$ spectra obtained with ALMA.  Each frequency channel of the contributing $3\1{mm}$ spectrum is weighted individually, so the average is an approximate value for the weight of each source.  

\begin{figure}[!h]
	\centering
	\includegraphics[width=\columnwidth]{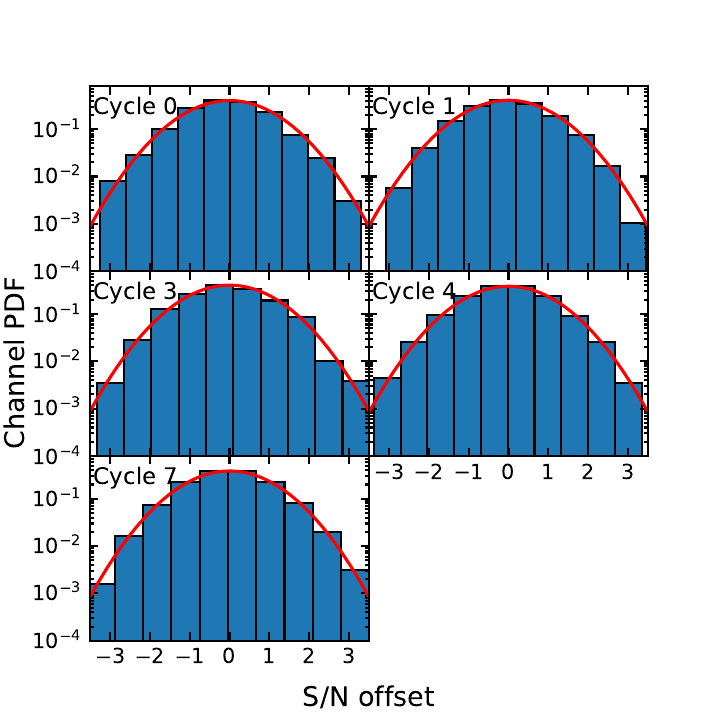}
	\caption{The signal to noise distribution of a line-free region of the stacking spectrum, shown as an illustration of the noise properties of the stacked spectrum.  The channels of the contributing spectra were randomized and stacked for a total of 3000 trials, split by ALMA Cycle.  For each ALMA Cycle, the resulting distribution is equivalent to a true Gaussian, with a width of $\sigma=1$. 
	\label{fig:noisemc}}
\end{figure}

The original stacked spectrum given in \citet{spilker14} was weighted by the inverse variance of noise, in order to optimize for signal to noise.  Because of ALMA's changing capabilities during our observations, we choose to weight by the inverse standard deviation instead, in order to mitigate the presence of outliers.  In order to test the statistical noise properties of the sample, we use the same procedure described in \citet{spilker14}.  That is, we select a line-free area of the spectrum ($513\1{GHz}$), shuffle the indices, scale and stack the spectra for a total of 3000 trials.  We then find the signal to noise ratio of the central bin for each trial, as was done in \citet{spilker14}.  
We find that the noise for each ALMA Cycle is well-described by Gaussian statistics, as shown in the resulting distributions in Fig.~\ref{fig:noisemc}.  

\begin{deluxetable}{lcclcc} 
\tabletypesize{\footnotesize}
\tablewidth{0pc}
\tablecolumns{6}
\tablehead{
\colhead{Source} & \colhead{Dust mass} & \colhead{$3\1{mm}$ noise} & \colhead{Source} & \colhead{Dust mass} & \colhead{$3\1{mm}$ noise}  \\ 
  & ($10^9$ M$_\odot$)& (mJy) & &  ($10^9$ M$_\odot$)& (mJy) \\ 
}
\startdata 
SPT0002-52 	&	 4.6 $_{2.1}^{6.1}$ 	&	 0.89 	&	 SPT0528-53 	&	 2.1 $_{1.2}^{4.0}$ 	&	 0.58 \\ 
SPT0020-51 	&	 7.2 $_{2.3}^{4.7}$ 	&	 0.65 	&	 SPT0529-54 	&	 23.3 $_{3.8}^{5.3}$ 	&	 1.19 \\ 
SPT0027-50 	&	 11.9 $_{2.0}^{2.9}$ 	&	 0.65 	&	 SPT0532-50 	&	 16.4 $_{3.2}^{5.3}$ 	&	 1.18 \\ 
SPT0054-41 	&	 7.2 $_{3.1}^{9.3}$ 	&	 0.56 	&	 SPT0544-40 	&	 6.5 $_{3.5}^{12.6}$ 	&	 0.81 \\ 
SPT0103-45 	&	 24.4 $_{4.4}^{6.0}$ 	&	 1.52 	&	 SPT0550-53 	&	 8.6 $_{4.3}^{12.9}$ 	&	 1.20 \\ 
SPT0106-64 	&	 8.3 $_{3.4}^{10.9}$ 	&	 0.70 	&	 SPT0551-50 	&	 9.2 $_{4.2}^{14.0}$ 	&	 1.19 \\ 
SPT0109-47 	&	 3.8 $_{3.0}^{4.5}$ 	&	 0.66 	&	 SPT0552-42 	&	 9.6 $_{4.1}^{14.6}$ 	&	 0.80 \\ 
SPT0112-55 	&	 0.9 $_{0.5}^{1.5}$ 	&	 0.29 	&	 SPT0553-50 	&	 3.5 $_{1.9}^{4.8}$ 	&	 0.78 \\ 
SPT0113-46 	&	 18.6 $_{3.3}^{6.0}$ 	&	 1.49 	&	 SPT0555-62 	&	 3.7 $_{1.9}^{4.9}$ 	&	 0.66 \\ 
SPT0125-47 	&	 11.8 $_{3.7}^{7.9}$ 	&	 1.47 	&	 SPT0604-64 	&	 19.5 $_{9.0}^{24.3}$ 	&	 0.67 \\ 
SPT0125-50 	&	 6.0 $_{1.3}^{1.7}$ 	&	 1.50 	&	 SPT0611-55 	&	 3.7 $_{1.6}^{4.6}$ 	&	 0.65 \\ 
SPT0136-63 	&	 7.8 $_{3.8}^{10.3}$ 	&	 0.71 	&	 SPT0625-58 	&	 19.1 $_{8.4}^{25.0}$ 	&	 0.65 \\ 
SPT0147-64 	&	 9.1 $_{4.6}^{15.6}$ 	&	 0.70 	&	 SPT0652-55 	&	 20.1 $_{7.9}^{22.9}$ 	&	 0.50 \\ 
SPT0150-59 	&	 11.2 $_{5.0}^{14.0}$ 	&	 0.70 	&	 SPT2031-51 	&	 12.2 $_{2.5}^{4.2}$ 	&	 0.58 \\ 
SPT0155-62 	&	 55.3 $_{23.6}^{68.4}$ 	&	 0.71 	&	 SPT2037-65 	&	 12.0 $_{4.8}^{15.7}$ 	&	 0.76 \\ 
SPT0202-61 	&	 7.8 $_{2.6}^{5.0}$ 	&	 0.71 	&	 SPT2048-55 	&	 7.1 $_{18.8}^{17.7}$ 	&	 0.75 \\ 
SPT0226-45 	&	 4.3 $_{1.9}^{4.6}$ 	&	 0.58 	&	 SPT2101-60 	&	 6.9 $_{3.2}^{7.6}$ 	&	 0.75 \\ 
SPT0243-49 	&	 20.1 $_{6.3}^{14.4}$ 	&	 1.77 	&	 SPT2103-60 	&	 8.6 $_{2.2}^{3.9}$ 	&	 1.76 \\ 
SPT0245-63 	&	 3.9 $_{1.1}^{2.6}$ 	&	 0.89 	&	 SPT2129-57 	&	 5.3 $_{2.4}^{7.3}$ 	&	 0.66 \\ 
SPT0300-46 	&	 8.2 $_{2.3}^{4.1}$ 	&	 1.65 	&	 SPT2132-58 	&	 5.3 $_{1.9}^{5.6}$ 	&	 1.72 \\ 
SPT0311-58 	&	 1.1 $_{0.7}^{2.1}$ 	&	 0.88 	&	 SPT2134-50 	&	 7.1 $_{2.1}^{4.8}$ 	&	 1.67 \\ 
SPT0314-44 	&	 14.5 $_{6.8}^{15.8}$ 	&	 0.56 	&	 SPT2146-55 	&	 5.4 $_{2.2}^{6.2}$ 	&	 1.70 \\ 
SPT0319-47 	&	 4.9 $_{1.6}^{3.9}$ 	&	 1.61 	&	 SPT2147-50 	&	 6.1 $_{1.4}^{1.9}$ 	&	 1.68 \\ 
SPT0345-47 	&	 7.2 $_{1.7}^{2.9}$ 	&	 1.58 	&	 SPT2152-40 	&	 7.3 $_{3.6}^{10.6}$ 	&	 0.52 \\ 
SPT0346-52 	&	 7.7 $_{1.5}^{2.5}$ 	&	 1.53 	&	 SPT2203-41 	&	 4.4 $_{2.0}^{6.3}$ 	&	 0.50 \\ 
SPT0348-62 	&	 2.7 $_{0.8}^{1.7}$ 	&	 0.88 	&	 SPT2232-61 	&	 7.3 $_{3.2}^{8.5}$ 	&	 0.65 \\ 
SPT0402-45 	&	 14.1 $_{6.4}^{16.6}$ 	&	 0.61 	&	 SPT2311-45 	&	 10.8 $_{4.8}^{13.6}$ 	&	 0.59 \\ 
SPT0403-58 	&	 3.6 $_{1.3}^{3.0}$ 	&	 0.60 	&	 SPT2311-54 	&	 3.3 $_{0.7}^{1.3}$ 	&	 0.93 \\ 
SPT0418-47 	&	 6.5 $_{1.6}^{2.3}$ 	&	 1.69 	&	 SPT2316-50 	&	 2.1 $_{1.2}^{4.0}$ 	&	 0.60 \\ 
SPT0425-40 	&	 3.6 $_{1.5}^{4.8}$ 	&	 0.61 	&	 SPT2319-55 	&	 2.7 $_{0.9}^{2.3}$ 	&	 0.89 \\ 
SPT0436-40 	&	 5.3 $_{2.4}^{7.2}$ 	&	 0.61 	&	 SPT2335-53 	&	 1.6 $_{0.8}^{2.4}$ 	&	 0.88 \\ 
SPT0441-46 	&	 6.1 $_{2.0}^{5.9}$ 	&	 1.54 	&	 SPT2340-59 	&	 2.3 $_{0.6}^{1.2}$ 	&	 0.20 \\ 
SPT0452-50 	&	 30.2 $_{4.9}^{7.3}$ 	&	 1.53 	&	 SPT2349-50 	&	 4.3 $_{1.2}^{3.0}$ 	&	 0.94 \\ 
SPT0457-49 	&	 4.2 $_{1.9}^{5.3}$ 	&	 0.23 	&	 SPT2349-52 	&	 1.0 $_{0.4}^{1.4}$ 	&	 0.57 \\ 
SPT0459-58 	&	 4.2 $_{1.3}^{2.8}$ 	&	 1.22 	&	 SPT2349-56 	&	 1.9 $_{0.4}^{0.6}$ 	&	 0.89 \\ 
SPT0459-59 	&	 5.7 $_{2.0}^{5.1}$ 	&	 1.23 	&	 SPT2351-57 	&	 2.1 $_{1.0}^{2.6}$ 	&	 0.90 \\ 
SPT0512-59 	&	 16.1 $_{7.4}^{18.9}$ 	&	 1.22 	&	 SPT2353-50 	&	 2.6 $_{1.3}^{3.6}$ 	&	 0.95 \\ 
SPT0516-59 	&	 2.6 $_{1.3}^{3.2}$ 	&	 0.66 	&	 SPT2354-58 	&	 6.0 $_{1.5}^{3.4}$ 	&	 0.91 \\ 
SPT0520-53 	&	 6.7 $_{3.3}^{8.5}$ 	&	 0.58 	&	 SPT2357-51 	&	 5.1 $_{0.9}^{1.2}$ 	&	 0.94 \\ 
\enddata 
\caption{Weights used in stacked spectrum.  The relative importance of each contributing source depends on two components: the apparent dust mass and the uncertainties from each frequency channel of the respective $3\1{mm}$ spectra, also shown in the top panel of Fig.~\ref{fig:noise}.  The apparent dust masses were calculated with the same photometry and method described in \citet{reuter20}, but using $\kappa_{850\1{\mu m}} /$ $\1{m^2}$ $\1{kg^{-1}} = 0.038$~\citep{draine03}.  Each frequency channel of the contributing $3\1{mm}$ spectrum is weighted individually, but the average values for each source are given above. \label{tab:weights} }
\end{deluxetable}

\section{Individual emission lines}
\label{ap:lines}
Tab.~\ref{tab:lines} gives a summary of lines detected in the stacked spectrum, along with upper limits for an assortment of ISM diagnostic lines.  The luminosities of these lines were obtained using the stacking procedure described in Sec.~\ref{sec:stackmethods} for ALMA $3\1{mm}$ data.  Lines with S/N$>3$ are shown in bold.  The luminosities are given with statistical error, which is derived from the noise of the contributing $3\1{mm}$ spectra.  These quantities are given along with the 16th and 84th percentiles of the luminosity distribution, as an illustration of the intrinsic scatter of the underlying population.  

\renewcommand{\theHtable}{B.\arabic{table}}
\renewcommand{\theHfigure}{B.\arabic{figure}}
\setcounter{table}{0}
\setcounter{figure}{0}
\renewcommand\thetable{\thesection.\arabic{table}}  
\renewcommand\thefigure{\thesection.\arabic{figure}} 

\vspace*{-\baselineskip}
\begin{deluxetable*}{lccccclccccc} 
\tabletypesize{\scriptsize}
\tablewidth{0pc}
\tablecolumns{12}
\tablehead{
\colhead{Line} & \colhead{$\nu_\mathrm{rest}$} & \colhead{$N_\mathrm{sources}$} & \colhead{$L'$} & \colhead{$dL'$} & \colhead{$_{16\mathrm{th}}^{84\mathrm{th}}$} &\colhead{Line} & \colhead{$\nu_\mathrm{rest}$} & \colhead{$N_\mathrm{sources}$} & \colhead{$L'$} & \colhead{$dL'$} & \colhead{$_{16\mathrm{th}}^{84\mathrm{th}}$} \\ 
 & \multicolumn{1}{c}{(GHz)} &  & \multicolumn{3}{c}{(K km/s pc$^2$)}  & & \multicolumn{1}{c}{(GHz)} & & \multicolumn{3}{c}{ (K km/s pc$^2$)}  \\
} %
\startdata 
CO(1-0)$^a$	&	115.2712	&	9	&	\textbf{234.60} &\textbf{9.53} &$_{-36.41}^{+715.39}$	&	CH 536	&	536.7614	&	37	&	\textbf{6.97} &\textbf{1.48} &$_{-9.96}^{+29.48}$	\\ 
CO(2-1)$^a$	&	230.5380	&	24	&	\textbf{261.90} &\textbf{6.53} &$_{-68.81}^{+752.38}$	&	OH 425	&	425.0363	&	30	&	-1.02 & 1.49 & $_{-20.41}^{+5.98}$	\\ 
CO(3-2)	&	345.7960	&	17	&	\textbf{200.40} &\textbf{4.70} &$_{-57.94}^{+542.29}$	&	OH 446	&	446.2910	&	36	&	\textbf{4.88} &\textbf{1.56} &$_{-9.88}^{+21.47}$	\\ 
CO(4-3)	&	461.0408	&	38	&	\textbf{202.20} &\textbf{4.60} &$_{-78.70}^{+597.13}$	&	CN($N=3-2$)	&	340.2478	&	15	&	\textbf{8.55} &\textbf{1.48} &$_{-6.89}^{+29.64}$	\\ 
CO(5-4)	&	576.2679	&	35	&	\textbf{188.10} &\textbf{5.51} &$_{-87.88}^{+563.67}$	&	CN($N=4-3$)	&	453.6067	&	35	&	3.65 & 1.24 & $_{-36.74}^{+77.05}$	\\ 
CO(6-5)	&	691.4731	&	12	&	\textbf{180.50} &\textbf{18.47} &$_{-67.97}^{+549.88}$	&	CN($N=5-4$)	&	566.9470	&	37	&	0.18 & 1.74 & $_{-13.28}^{+20.89}$	\\ 
$^{13}$CO(3-2)	&	330.5880	&	14	&	\textbf{9.96} &\textbf{1.85} &$_{-10.77}^{+30.14}$	&	CN($N=6-5$)	&	680.2641	&	12	&	-8.14 & 4.08 & $_{-14.15}^{+13.63}$	\\ 
$^{13}$CO(4-3)	&	440.7652	&	32	&	\textbf{6.66} &\textbf{1.54} &$_{-13.85}^{+26.22}$	&	SiO(7-6)	&	303.9270	&	9	&	-2.26 & 2.14 & $_{-20.29}^{+0.06}$	\\ 
$^{13}$CO(5-4)	&	550.9263	&	36	&	\textbf{6.80} &\textbf{1.72} &$_{-11.32}^{+28.74}$	&	SiO(8-7)	&	347.3306	&	16	&	-1.40 & 1.57 & $_{-11.14}^{+2.44}$	\\ 
$^{13}$CO(6-5)	&	661.0673	&	19	&	5.41 & 3.16 & $_{-23.86}^{+34.40}$	&	SiO(9-8)	&	390.7284	&	26	&	-1.92 & 1.46 & $_{-12.33}^{+2.83}$	\\ 
C$^{18}$O(3-2)	&	329.3305	&	15	&	3.44 & 1.83 & $_{-9.10}^{+15.32}$	&	SiO(10-9)	&	434.1196	&	32	&	-0.73 & 1.50 & $_{-12.16}^{+14.67}$	\\ 
C$^{18}$O(4-3)	&	439.0888	&	30	&	2.26 & 1.70 & $_{-17.46}^{+16.20}$	&	SiO(11-10)	&	477.5031	&	34	&	-0.16 & 1.51 & $_{-7.06}^{+14.47}$	\\ 
C$^{18}$O(5-4)	&	548.8310	&	36	&	2.10 & 1.54 & $_{-19.03}^{+21.88}$	&	SiO(12-11)	&	520.8782	&	36	&	0.14 & 1.53 & $_{-11.95}^{+16.64}$	\\ 
C$^{18}$O(6-5)	&	658.5533	&	19	&	2.02 & 3.19 & $_{-15.48}^{+21.71}$	&	SiO(13-12)	&	564.2440	&	36	&	-2.29 & 1.84 & $_{-17.14}^{+5.46}$	\\ 
$[$CI$]$(1-0)	&	492.1606	&	38	&	\textbf{51.85} &\textbf{1.51} &$_{-20.95}^{+169.53}$	&	SiO(14-13)	&	607.5994	&	27	&	1.08 & 2.12 & $_{-19.07}^{+32.47}$	\\ 
HCN(3-2)	&	265.8864	&	3	&	-2.76 & 5.49 & $_{-17.88}^{+6.44}$	&	SiO(15-14)	&	650.9436	&	19	&	-3.40 & 3.05 & $_{-20.45}^{+11.99}$	\\ 
HCN(4-3)	&	354.5055	&	18	&	\textbf{10.74} &\textbf{1.46} &$_{-10.83}^{+30.03}$	&	SiO(16-15)	&	694.2754	&	11	&	-0.81 & 6.21 & $_{-26.69}^{+16.00}$	\\ 
HCN(5-4)	&	443.1161	&	32	&	2.78 & 1.52 & $_{-14.72}^{+18.93}$	&	SiO(17-16)	&	737.5939	&	7	&	-9.79 & 7.04 & $_{-17.87}^{+-10.89}$	\\ 
HCN(6-5)	&	531.7164	&	37	&	3.24 & 1.50 & $_{-11.70}^{+29.67}$	&	CS(6-5)	&	293.9122	&	7	&	-0.60 & 2.33 & $_{-4.32}^{+3.19}$	\\ 
HCN(7-6)	&	620.3040	&	24	&	2.83 & 3.00 & $_{-20.20}^{+19.66}$	&	CS(7-6)	&	342.8830	&	16	&	0.44 & 1.48 & $_{-6.17}^{+5.73}$	\\ 
HCN(8-7)	&	708.8770	&	10	&	8.63 & 7.03 & $_{-36.55}^{+48.19}$	&	CS(8-7)	&	391.8470	&	26	&	0.90 & 1.42 & $_{-8.39}^{+9.78}$	\\ 
HNC(3-2)	&	271.9811	&	3	&	0.40 & 5.17 & $_{-10.36}^{+13.37}$	&	CS(10-9)	&	489.7510	&	38	&	1.52 & 1.47 & $_{-12.88}^{+18.70}$	\\ 
HNC(4-3)	&	362.6303	&	20	&	\textbf{6.41} &\textbf{1.63} &$_{-9.47}^{+34.79}$	&	CS(11-10)	&	538.6888	&	37	&	-1.42 & 1.47 & $_{-17.95}^{+8.77}$	\\ 
HNC(5-4)	&	453.2699	&	35	&	\textbf{5.23} &\textbf{1.56} &$_{-22.82}^{+41.21}$	&	CS(12-11)	&	587.6162	&	32	&	3.80 & 1.81 & $_{-8.42}^{+26.74}$	\\ 
HNC(6-5)	&	543.8976	&	38	&	1.11 & 1.52 & $_{-13.66}^{+17.30}$	&	CS(13-12)	&	636.5318	&	20	&	0.73 & 3.26 & $_{-18.06}^{+25.13}$	\\ 
HNC(7-6)	&	634.5108	&	21	&	0.68 & 3.30 & $_{-10.10}^{+18.61}$	&	CS(14-13)	&	685.4348	&	12	&	-12.21 & 4.66 & $_{-27.14}^{+-9.62}$	\\ 
HNC(8-7)	&	725.1073	&	8	&	0.86 & 7.09 & $_{-10.62}^{+22.78}$	&	CS(15-14)	&	734.3240	&	7	&	-7.10 & 6.91 & $_{-26.66}^{+-2.15}$	\\ 
HCO$^+$(3-2)	&	267.5576	&	3	&	7.97 & 5.41 & $_{-2.91}^{+42.52}$	&	NH$_3$(1$_{0}$-0$_{0}$)	&	572.4982	&	35	&	0.41 & 1.77 & $_{-21.37}^{+20.01}$	\\ 
HCO$^+$(4-3)	&	356.7342	&	19	&	\textbf{9.27} &\textbf{1.48} &$_{-9.96}^{+33.25}$	&	N$_2$H$^+$(3-2)	&	279.5117	&	3	&	0.22 & 4.59 & $_{-13.75}^{+9.24}$	\\ 
HCO$^+$(5-4)	&	445.9029	&	34	&	3.36 & 1.54 & $_{-7.61}^{+16.27}$	&	N$_2$H$^+$(4-3)	&	372.6725	&	22	&	0.27 & 1.44 & $_{-10.98}^{+4.83}$	\\ 
HCO$^+$(6-5)	&	535.0616	&	38	&	3.24 & 1.48 & $_{-11.90}^{+29.54}$	&	N$_2$H$^+$(5-4)	&	465.8250	&	36	&	1.12 & 1.56 & $_{-8.35}^{+13.46}$	\\ 
HCO$^+$(7-6)	&	624.2085	&	24	&	3.72 & 3.13 & $_{-16.55}^{+23.81}$	&	N$_2$H$^+$(6-5)	&	558.9667	&	34	&	2.60 & 1.73 & $_{-9.96}^{+27.52}$	\\ 
HCO$^+$(8-7)	&	713.3414	&	9	&	3.06 & 7.48 & $_{-29.43}^{+14.60}$	&	N$_2$H$^+$(7-6)	&	652.0959	&	19	&	1.19 & 3.06 & $_{-10.83}^{+21.62}$	\\ 
H$_2$O(5$_{1,5}$-4$_{2,2}$)	&	325.1529	&	13	&	-0.78 & 1.78 & $_{-5.28}^{+7.38}$	&	N$_2$H$^+$(8-7)	&	745.2103	&	7	&	5.86 & 8.18 & $_{-38.25}^{+31.59}$	\\ 
H$_2$O(4$_{1,4}$-3$_{2,1}$)	&	380.1974	&	25	&	\textbf{4.04} &\textbf{1.32} &$_{-9.88}^{+21.01}$	&	CCH(3-2)	&	262.0042	&	3	&	-2.26 & 5.15 & $_{-25.94}^{+-3.88}$	\\ 
H$_2$O(4$_{2,3}$-3$_{3,0}$)	&	448.0011	&	35	&	-1.63 & 1.64 & $_{-14.08}^{+8.02}$	&	CCH(4-3)	&	349.3387	&	17	&	3.10 & 1.56 & $_{-4.17}^{+18.84}$	\\ 
H$_2$O(5$_{3,3}$-4$_{4,0}$)	&	474.6891	&	36	&	2.18 & 1.59 & $_{-18.20}^{+19.10}$	&	CCH(5-4)	&	436.6604	&	30	&	-0.30 & 1.54 & $_{-14.42}^{+15.39}$	\\ 
H$_2$O(1$_{1,0}$-1$_{0,1}$)	&	556.9360	&	37	&	\textbf{-14.69} &\textbf{1.72} &$_{-23.85}^{+-6.43}$	&	CCH(6-5)	&	523.9704	&	37	&	2.34 & 1.51 & $_{-18.15}^{+22.26}$	\\ 
H$_2$O(2$_{1,1}$-2$_{0,2}$)	&	752.0331	&	7	&	\textbf{38.95} &\textbf{7.44} &$_{-32.87}^{+152.72}$	&	CCH(7-6)	&	611.2650	&	24	&	1.12 & 2.21 & $_{-22.91}^{+23.19}$	\\ 
p-H$_2$O$^+$ 604	&	604.6786	&	29	&	-0.69 & 2.11 & $_{-14.88}^{+40.00}$	&	CCH(8-7)	&	698.5416	&	11	&	5.55 & 6.62 & $_{-15.85}^{+33.83}$	\\ 
p-H$_2$O$^+$ 607	&	607.2273	&	27	&	-2.19 & 2.07 & $_{-28.85}^{+10.62}$	&	H21$\alpha$	&	662.4042	&	17	&	-1.38 & 3.18 & $_{-15.59}^{+19.89}$	\\ 
p-H$_2$O$^+$ 631	&	631.7241	&	20	&	0.55 & 3.29 & $_{-6.79}^{+21.55}$	&	H23$\alpha$	&	507.1755	&	37	&	0.62 & 1.74 & $_{-10.86}^{+11.03}$	\\ 
p-H$_2$O$^+$ 634	&	634.2729	&	20	&	2.34 & 3.24 & $_{-14.31}^{+16.55}$	&	H24$\alpha$	&	447.5403	&	35	&	0.42 & 1.66 & $_{-18.04}^{+15.88}$	\\ 
o-H$_2$O$^+$ 721	&	721.9274	&	9	&	3.58 & 6.84 & $_{-27.63}^{+28.32}$	&	H25$\alpha$	&	396.9008	&	25	&	0.34 & 1.39 & $_{-4.97}^{+23.67}$	\\ 
o-H$_2$O$^+$ 742	&	742.1090	&	7	&	15.96 & 7.32 & $_{-32.92}^{+47.04}$	&	H26$\alpha$	&	353.6227	&	18	&	-1.45 & 1.48 & $_{-5.20}^{+11.46}$	\\ 
o-H$_2$O$^+$ 746	&	746.5417	&	7	&	15.01 & 7.02 & $_{-26.62}^{+57.77}$	&	H27$\alpha$	&	316.4154	&	11	&	-3.00 & 2.03 & $_{-5.84}^{+1.85}$	\\ 
o-H$_2$O$^+$ 761	&	761.8188	&	5	&	11.03 & 8.10 & $_{-8.50}^{+51.68}$	&	H28$\alpha$	&	284.2506	&	5	&	2.92 & 3.91 & $_{-8.23}^{+24.62}$	\\ 
CH 532	&	532.7239	&	37	&	\textbf{11.49} &\textbf{1.52} &$_{-9.34}^{+51.71}$	&		&		&		&	& &	\\ 
\enddata 
\caption{The luminosities were obtained using the stacking procedure described in Sec.~\ref{sec:stacked_spec} for ALMA $3\1{mm}$ data.  Lines with S/N$>3$ are shown in bold.  The luminosities are given with statistical error, along with the 16th and 84th percentiles of the intrinsic scatter.  For transitions with fine or hyperfine structure, only the main transition is listed and the line may be referred to by frequency for clarity.  \label{tab:lines} }
\tablenotetext{a}{Derived from low-J $^{12}$CO observations obtained from ATCA.  While 17 sources have been presented in \citet{aravena16}, the remainder will be published in a future work.}
\end{deluxetable*}

\end{document}